\documentclass[journal]{IEEEtran}
\usepackage{cite}
\usepackage{color,soul}
\usepackage{gensymb}
\usepackage{dsfont}
 \usepackage{bbm}

\ifCLASSINFOpdf
  \usepackage[pdftex]{graphicx}
  \graphicspath{{../pdf/}{../jpeg/}}
  \DeclareGraphicsExtensions{.pdf,.jpeg,.png}
\else
  \usepackage[dvips]{graphicx}
  \graphicspath{{../eps/}}
  \DeclareGraphicsExtensions{.eps}
\fi
\usepackage[cmex10]{amsmath}

\usepackage{amssymb}
{}

\usepackage{euscript}

\usepackage{multirow}
\usepackage{algorithm}
\usepackage{algpseudocode}
\usepackage{diagbox}
\usepackage{physics}
\usepackage{bm}
\usepackage{tikz}
\usepackage{comment}

\usepackage{amsmath,lipsum}
\usepackage{cuted}

\usepackage{xcolor,colortbl}
\definecolor{myred}{RGB}{205 38 38}
 \usepackage{bbm}
\definecolor{myblue}{RGB}{105,89,205}
\definecolor{myorange}{RGB}{238,92,66}
\definecolor{mygray}{RGB}{205,201,201}

\ifCLASSOPTIONcompsoc
  \usepackage[caption=false,font=normalsize,labelfont=sf,textfont=sf]{subfig}
\else
  \usepackage[caption=false,font=footnotesize]{subfig}
\fi
\usepackage{url}
\allowdisplaybreaks[4]

\begin{document}
%
\title{A Two-Level Simulation-Assisted Sequential Distribution System Restoration Model With Frequency Dynamics Constraints}

%
%
%

\author{
         Qianzhi Zhang,~\IEEEmembership{Student Member,~IEEE,}
         Zixiao Ma,~\IEEEmembership{Student Member,~IEEE,}
         Yongli Zhu,~\IEEEmembership{Member,~IEEE,}
         
         and Zhaoyu Wang,~\IEEEmembership{Senior Member,~IEEE.}
\thanks{
This work was supported in part by the U.S. Department of Energy Wind Energy Technologies Office under Grant DE-EE00008956 (\emph{Corresponding author: Zhaoyu Wang}).

The authors are with the Department of
Electrical and Computer Engineering, Iowa State University, Ames,
IA 50011 USA (e-mail: qianzhi@iastate.edu; zma@iastate.edu; yongliz@iastate.edu; wzy@iastate.edu).}}

\maketitle

\begin{abstract}
This paper proposes a service restoration model for unbalanced distribution systems and inverter-dominated microgrids (MGs), in which frequency dynamics constraints are developed to optimize the amount of load restoration and guarantee the dynamic performance of system frequency response during the restoration process. After extreme events, the damaged distribution systems can be sectionalized into several isolated MGs to restore critical loads and tripped non-black start distributed generations (DGs) by black start DGs. However, the high penetration of inverter-based DGs reduces the system inertia, which results in low-inertia issues and large frequency fluctuation during the restoration process. To address this challenge, we propose a two-level simulation-assisted sequential service restoration model, which includes a mixed integer linear programming (MILP)-based optimization model and a transient simulation model. The proposed MILP model explicitly incorporates the frequency response into constraints, by interfacing with transient simulation of inverter-dominated MGs. Numerical results on a modified IEEE 123-bus system have validated that the frequency dynamic performance of the proposed service restoration model are indeed improved.      
\end{abstract}


\section*{Nomenclature}
\addcontentsline{toc}{section}{Nomenclature}
\begin{IEEEdescription}[\IEEEusemathlabelsep\IEEEsetlabelwidth{$V_1,V_2,V_3$}]
\item[\textbf{Sets}]
\item[$\Omega_{\rm BK}$] Set of bus blocks.
\item[$\Omega_{\rm G}$] Set of generators.
\item[$\Omega_{\rm BS}$] Set of generators with black start capability.
\item[$\Omega_{\rm NBS}$] Set of generators without black start capability.
\item[$\Omega_{\rm K}$] Set of distribution lines.
\item[$\Omega_{\rm SW_K}$] Set of switchable lines.
\item[$\Omega_{\rm NSW_K}$] Set of non-switchable lines.
\item[$\Omega_{\rm L}$] Set of loads.
\item[$\Omega_{\rm SW_L}$] Set of switchable loads.
\item[$\Omega_{\rm NSW_L}$] Set of non-switchable loads.
\item[$\Omega_{\phi}$] Set of phases.

\item[\textbf{Indices}]
\item[$BK$] Index of bus block.
\item[$k$] Index of line.
\item[$i,j$] Index of bus.
\item[$t$] Index of time instant.
\item[$\phi$] Index of three-phase $\phi_a,\phi_b,\phi_c$.

\item[\textbf{Parameters}]
\item[$a_{\phi}$] Approximate relative phase unbalance.
\item[$D_{\rm P},D_{\rm Q}$] $P-\omega$ and $Q-V$ droop gains.
\item[$f_0$] Nominal steady-state frequency.
\item[$f^{\rm min}$] Minimum allowable frequency during the transient simulation.
\item[$M$] Big-M number.
\item[$P_{i}^{\rm G,M},Q_{i}^{\rm G,M}$] Active and reactive power output maximum limits of generator at bus $i$.
\item[$P_{k}^{\rm K,M},Q_{k}^{\rm K,M}$] Active and reactive power flow maximum limits of line $k$.
\item[$p_{k,\phi}$] Phase identifier of line $k$.
\item[$R,L$] Aggregate resistance and inductance of connections from the inverter terminal's point review.
\item[$\hat{R}_{k},\hat{X}_{k}$] Matrices of resistance and reactance of line $k$.
\item[$T$] Length of rolling horizon.
\item[$U_{i}^{\rm m},U_{i}^{\rm M}$] Minimum and maximum limit for squared nodal voltage magnitude of bus $i$.
\item[$V_{\rm bus}$] Bus voltage.
\item[$Z_{k},\hat{Z}_{k}$] Matrices of original impedance and equivalent impedance of line $k$.
\item[$\alpha$] Hyper-parameter in frequency dynamics constraints.
\item[$\Delta f^{\rm max}$] User-defined maximum allowable frequency drop limit.
\item[$\Delta f^{\rm meas}$] Measured maximum transient frequency drop.
\item[$w_i^{\rm L}$] Priority weight factor for load of bus $i$.
\item[$\omega_{\rm c}$] Cut-off frequency of the low pass filter.
\item[$\omega_{\rm set},V_{\rm set}$] Set points of frequency and voltage controllers.
\item[$\omega_0$] Nominal angular frequency.

\item[\textbf{Variables}]
\item[$f^{\rm nadir}$] Frequency nadir during the transient simulation.
\item[$I_{\rm d},I_{\rm q}$] $dq$-axis current.
\item[$P,Q$] Filtered terminal output active and reactive power.
\item[$P^{\rm L},Q^{\rm L}$] Restored active and reactive loads.
\item[$P_{i,\phi,t}^{\rm G}$] Three-phase active power output of generator at bus $i$, phase $\phi$, time $t$.
\item[$P_{i,t}^{\rm G,MLS}$] Maximum load step at bus $i$, time $t$.
\item[$P_{k,\phi,t}^{\rm K}$] Three-phase active power flow of line $k$, phase $\phi$, time $t$.
\item[$P_{i,\phi,t}^{\rm L}$] Restored active load at bus $i$, phase $\phi$, time $t$.
\item[$Q_{i,\phi,t}^{\rm G}$] Three-phase reactive power output of generator at bus $i$, phase $\phi$, time $t$.
\item[$Q_{k,\phi,t}^{\rm K}$] Three-phase reactive power flow of line $k$, phase $\phi$, time $t$.
\item[$U_{i,\phi,t}$] Squared of three-phase voltage magnitude.
\item[$V$] Output voltage of the inverter.
\item[$x_{i,t}^{\rm B}$] Binary energizing status of bus, if $x_{i,t}^{\rm B}=1$ then the bus $i$ is energized at time $t$.
\item[$x_{B,t}^{\rm BK}$] Binary energizing status of bus block, if $x_{B,t}^{\rm BK}=1$ then the bus block $B$ is energized at time $t$.
\item[$x_{i,t}^{\rm G}$] Binary switch on/off status of grid-following generator, if $x_{i,t}^{\rm G}=1$ then the grid-following generator at bus $i$ is switched on at time $t$.
\item[$x_{k,t}^{\rm K}$] Binary connection status of line, if $x_{k,t}^{\rm K}=1$ then the line $k$ is connected at time $t$.
\item[$x_{i,t}^{\rm L}$] Binary restoration status of load, if $x_{i,t}^{\rm L}=1$ then the load $i$ is restored at time $t$.
\item[$\Delta P_{i,t-1}^{G,MLS}$] Change of the maximum load step.
\item[$\theta$] Output phase angle of the inverter.
\item[$\omega$] Output angular frequency of the inverter.                                         
\end{IEEEdescription}

%
\IEEEpeerreviewmaketitle

\section{Introduction}
\IEEEPARstart{E}XTREME events can cause severe damages to power distribution systems \cite{WH_report}, e.g. substation disconnection, line outage, generator tripping, load shedding, and consequently large-scale system blackouts \cite{salman2015evaluating}. During the network and service restoration, in order to isolate faults and restore critical loads, a distribution system can be sectionalized into several isolated microgirds (MGs) \cite{MG_BS}. Through the MG formation, buses, lines and loads in outage areas can be locally energized by distributed generations (DGs), where more outage areas could be restored and the number of switching operations could be minimized \cite{Restoration_1,Restoration_2,Restoration_3,Restoration_4,Restoration_5,Restoration_6}. In \cite{Restoration_1}, the self-healing mode of MGs is considered to provide reliable power supply for critical loads and restore the outage areas. In \cite{Restoration_2}, a networked MGs-aided approach is developed for service restoration, which considers both dispatchable and non-dispatchable DGs. In \cite{Restoration_3} and \cite{Restoration_4}, the service restoration problem is formulated as a mixed integer linear programming (MILP) to maximize the critical loads to be restored while satisfying constraints for MG formation and remotely controlled devices. In \cite{Restoration_5}, the formation of adaptive multiple MGs is developed as part of the critical service restoration strategy. In \cite{Restoration_6}, a sequential service restoration framework is proposed to generate restoration solutions for MGs in the event of large-scale power outages. However, the previous methods mainly use the conventional synchronous generators as the black start units, and only consider steady-state constraints in the service restoration models, which have limitations in the following aspects:

(1) \textit{An inverter-dominated MG can have low-inertia:} With the increasing penetration of inverter-based DGs (IBDGs) in distribution systems, such as distributed wind and photovoltaics (PVs) generations, the system inertia becomes lower \cite{Low_inertia_1,PLL_Gu}. When sudden changes happen, such as DG output changing, load reconnecting, and line switching, the dynamic frequency performance of such low-inertia distribution systems can deteriorate \cite{Low_inertia_2}. This issue becomes even worse when restoring low-inertia inverter-dominated MGs. Without considering frequency dynamics constraints, the load and service restoration decisions may not be implemented in practice.

(2) \textit{Frequency responses need to be considered:} Previous studies \cite{Restoration_stability_1,Restoration_stability_2,Restoration_stability_3,Restoration_stability_5} have considered the impact of disturbances on frequency responses in the service restoration problem using different approaches. In \cite{Restoration_stability_1}, the amount of load restored by DGs is limited by a fixed frequency response rate and maximum allowable frequency deviation. However, because the frequency response rate is pre-determined in an off-line manner, the impacts of significant load restoration, topology change, and load variations may not be fully captured by the off-line model. In \cite{Restoration_stability_2}, the stability and security constraints are incorporated into the restoration model. However, this model has to be solved by meta-heuristic methods due to the nonlinearity of the stability constraints, which may lead to large optimality gaps. In \cite{Restoration_stability_3}, even though the transient simulation results of voltage and frequency are considered to evaluate the potential MG restoration paths in an online manner, it adopts a relatively complicated four-stage procedure to obtain the optimal restoration path. In \cite{Restoration_stability_5}, a control strategy of real-time frequency regulation for network reconfiguration is developed, nonetheless, it is not co-optimized with the switch operations.

(3) \textit{Grid-forming IBDGs need to be considered:} In previous studies on optimal service restoration, IBDGs are usually modeled as grid-following sources (i.e., PQ sources) to simply supply active and reactive power based on the control commands. However, during the service restoration after a network blackout and loss of connection to the upstream feeder, a grid-forming IBDG will be needed to setup voltage and frequency references for the blackout network \cite{Inverters_model_2}. During outages, the grid-following IBDGs will be switched off. After outages, the grid-forming IBDGs have the black start capability, which can restore loads after the faults are isolated. Because IBDGs are connected with power electronics converters and have no rotating mass, there is no conventional concept of ``inertia" for IBDGs. Thus, control techniques such as droop control \cite{Inverters_model_1,Inverters_model_3} and virtual synchronous generator (VSG) control \cite{VSG_1,VSG_2} are usually adopted to emulate the inertia property in IBDGs. 

To alleviate the frequency fluctuations caused by service restoration, we establish a MILP-based optimization model with frequency dynamics constraints for sequential service restoration to generate sequential actions for remotely controlled switches, restoration status for buses, lines, loads, operation actions for grid-forming and grid-following IBDGs, which interacts with the transient simulation of inverter-dominated MGs. Inspired by recent advances in simulation-assisted methods \cite{Restoration_stability_3,YL} and to incorporate the frequency dynamics constraints \textit{explicitly} in the optimization formulation, we associate the frequency nadir of the transient simulation with respect to the maximum load that a MG can restore. Although some previous works have considered the transient simulation as well in finding the optimal restoration solution, they either adopts a heuristic framework, or merely using the transient simulation to validate the feasibility of the obtained restoration solution after solving an optimization problem. By contrast, the proposed two-level simulation-assisted restoration model directly incorporates the transient simulation module on top of a strict MILP optimization problem via explicit constraints, thus its solving process is more tractable and straightforward. 

The main contribution of this paper is two-folded:
\begin{itemize}
\item We develop a two-level simulation-assisted sequential service restoration model within a rolling horizon framework, which combines a MILP-based optimization level of service restoration and a transient simulation level of inverter-dominated MGs.
\item Frequency dynamics constraints are developed and explicitly incorporated in the optimization model, to associate the simulated frequency responses with the decision variables of maximum load step at each stage. These constraints help restrict the system frequency drop during the transient periods of restoration. Thus, the generated restoration solution can be more secure and practical.
\end{itemize}

The reminder of the paper is organized as follows: Section \ref{sec:framework} presents the overall framework of the proposed service restoration model. Section \ref{sec:Level_I} introduces frequency dynamics constrained MILP-based sequential service restoration. Section \ref{sec:Level_II} describes transient simulation of inverter-dominated MGs. Numerical results and conclusions are given in Section \ref{sec:Results} and Section \ref{sec:Con}, respectively. 

\section{Overview of the Proposed service restoration Model}\label{sec:framework}
\begin{figure}
	\vspace{-0pt} 
	\vspace{-0pt}
	\centering
	\includegraphics[width=1.0\linewidth]{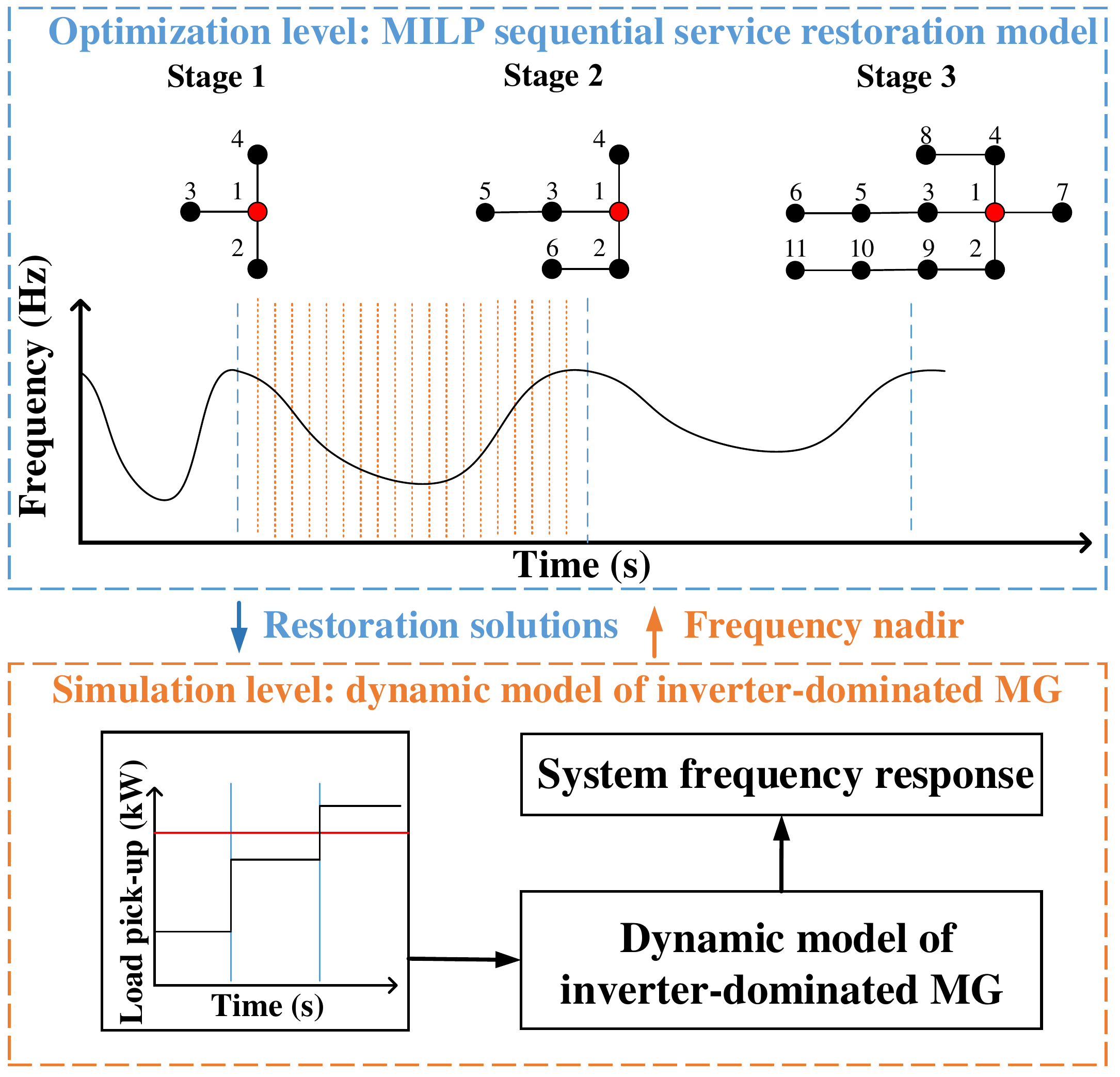}
	\vspace{-0pt} 
	\caption{The overall framework of the proposed service restoration model with optimization level and simulation level.}
	\centering
	\label{framework}
    \vspace{-0pt} 
\end{figure}

The general framework of the proposed two-level simulation-assisted service restoration is shown in Fig. \ref{framework}, including an optimization level of MILP-based sequential service restoration model and a transient simulation level of $7$th-order electromagnetic inverter-dominated MG dynamic model. After outages, the fault-affected areas of the distribution system will be isolated. Consequently, each isolated sub-network can be considered as a MG \cite{Restoration_stability_4}, which can be formed by the voltage and frequency supports from the grid-forming IBDGs, and active and reactive power supplies from the grid-following IBDGs. In the proposed optimization level, each MG will determine its restoration solutions, including optimal service restoration status of loads, optimal operation of remotely controlled switches and optimal active and reactive power dispatches of IBDGs. To prevent large frequency fluctuation due to a large load restoration, the maximum restorable load for a given period is limited by the proposed frequency dynamics constraints. In this way, the whole restoration process is divided into multiple stages. As shown in Fig. \ref{framework}, the information exchanged between the optimization level and the simulation level are the restoration solution (obtained from optimization) and MG system frequency nadir value (obtained from transient simulation): at each restoration stage, the optimization level will obtain and send the optimal restoration solution to the simulation level; then, after receiving the restoration solution, the simulation level will begin to run transient simulation by the proposed dynamic model of each inverter-dominated MG, and send the frequency nadir value to the optimization level for next restoration stage. 

\begin{figure}
	\vspace{-0pt} 
	\vspace{-0pt}
	\centering
	\includegraphics[width=0.85\linewidth]{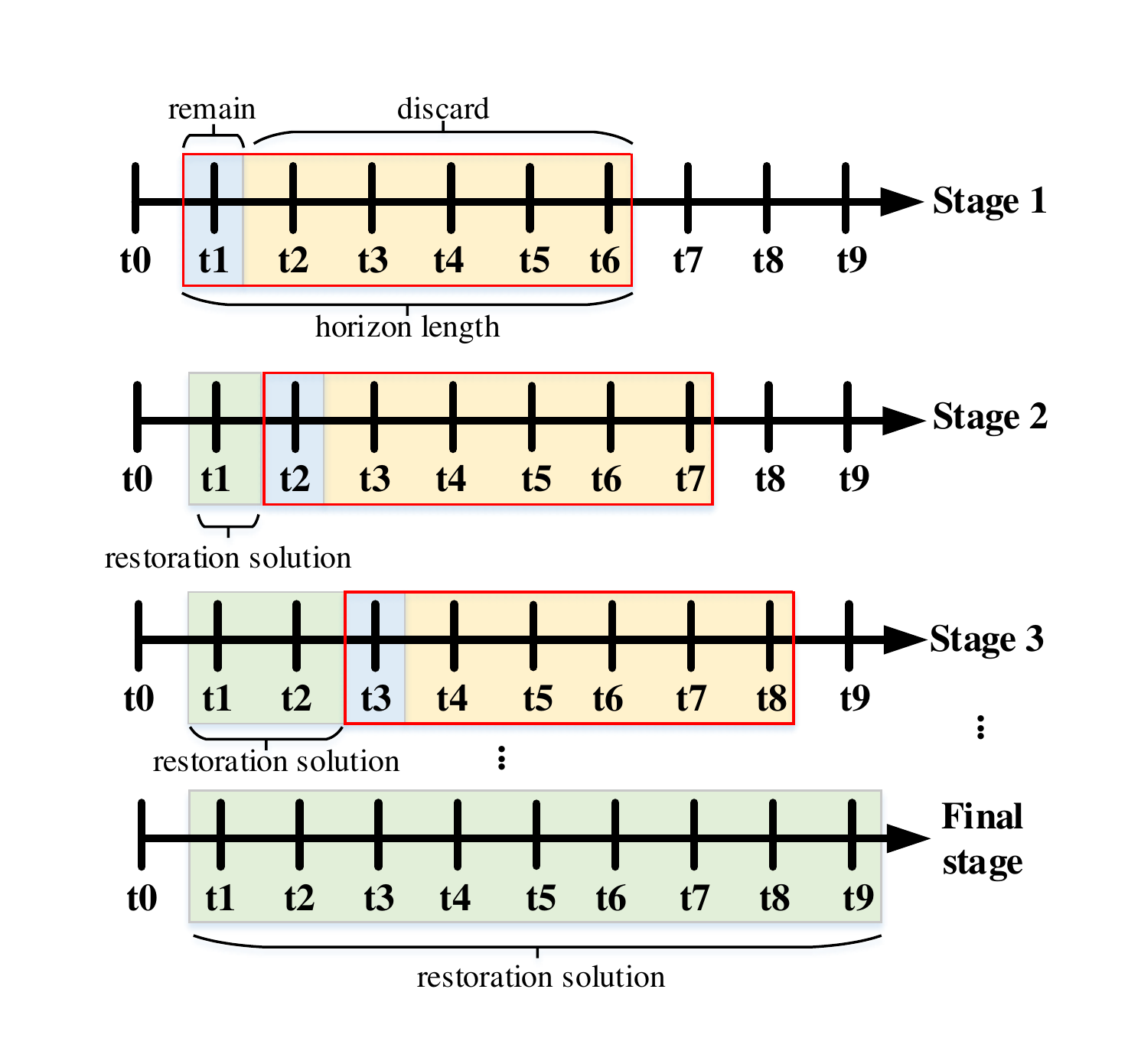}
	\vspace{-0pt} 
	\caption{Implementation of rolling-horizon in the proposed restoration model.}
	\centering
	\label{rolling_hor}
    \vspace{-0pt} 
\end{figure}

To accurately reflect the dynamic frequency-supporting capacities of grid-forming IBDGs during the service restoration process, a rolling-horizon framework is implemented in the proposed service restoration model, as shown in Fig. \ref{rolling_hor}. More specifically, we \textit{repeatedly} run the MILP-based sequential service restoration model by incorporating the network configuration from the preceding stage as the initial condition, and then feedback the frequency nadir value from the transient simulation to the frequency dynamics constraints. For each stage: (1) the horizon length will be fixed; (2) then only the restoration solution of first horizon of the current stage is retained and transferred to the simulation level, while the remaining horizons are discarded; (3) this process will keep going until the maximum restored load is reached in each MG. More details about the principles of rolling horizon can be found in \cite{MPC}.    

\section{Frequency Dynamics Constrained Service Restoration}\label{sec:Level_I}
This section presents the mathematical formulation for coordinating remotely controlled switches, grid-forming and grid-following IBDGs, and the sequential restoration status of buses, lines and loads. Here, we consider a unbalanced three-phase radial distribution system. The three-phase $\phi_a,\phi_b,\phi_c$ are simplified as $\phi$. Define the set $\Omega_{\rm L}=\Omega_{\rm SW_L}\cup\Omega_{\rm NSW_L}$, where $\Omega_{\rm SW_L}$ and $\Omega_{\rm NSW_L}$ represent the set of switchable load and the set of non-switchable loads, respectively. Define the set $\Omega_{\rm G}=\Omega_{\rm BS}\cup\Omega_{\rm NBS}$, where $\Omega_{\rm BS}$ and $\Omega_{\rm NBS}$ represent the set of grid-forming IBDGs with black start capability and the set of grid-following IBDGs without black start capability, respectively. Define the set $\Omega_{\rm K}=\Omega_{\rm SW_K}\cup\Omega_{\rm NSW_K}$, where $\Omega_{\rm SW}$ and $\Omega_{\rm NSW}$ represent the set of switchable lines and the set of non-switchable lines, respectively. Define $\Omega_{\rm BK}$ as the set of bus blocks, where bus block \cite{Restoration_6} is a group of buses interconnected by non-switchable lines and those bus blocks are interconnected by switchable lines. It is assumed that bus block can be energized by grid-forming IBDGs. By forcing the related binary variables of faulted lines to be zeros, each faulted area remains isolated during the restoration process. 

\subsection{MILP-based Sequential Service Restoration Formulation}
The objective function \eqref{eq3_obj_1} aims to maximize the total restored loads with priority factor $w_i^L$ over a rolling horizon $[t,t+T]$ as shown below: 
\begin{equation}
\max \sum_{t\in[t,t+T]}\sum_{i\in \Omega_L}\sum_{\phi\in \Omega_\phi}(w_i^{\rm L} x_{i,t}^{\rm L} P_{i,\phi,t}^{\rm L})\label{eq3_obj_1}
\end{equation}
where $P_{i,\phi,t}^{\rm L}$ and $x_{i,t}^{\rm L}$ are the restored load and restoration status of load at $t$. If the load demand $P_{i,\phi,t}^{\rm L}$ is restored, then $x_{i,t}^{\rm L}=1$. $T$ is horizon length in the rolling horizon optimization problem. In this work, the amount of restored load is also bounded by frequency dynamics constraints with respect to frequency response and maximum load step. More details of frequency dynamics constraints are discussed in Section \ref{sec:MLS}. 

Constraints \eqref{eq3_bal_1}-\eqref{eq3_U_2} are defined by the unbalanced three-phase version of linearized DistFlow model \cite{unbal,QZ_CVR} in each formed MG during the service restoration process. Constraints \eqref{eq3_bal_1} and \eqref{eq3_bal_2} are the nodal active and reactive power balance constraints, where $P_{k,\phi,t}^{\rm K}$ and $Q_{k,\phi,t}^{\rm K}$ are the active and reactive power flows along line $k$, and $P^{\rm G}_{i,\phi,t}$ and $Q^{\rm G}_{i,\phi,t}$ are the power outputs of the generators. Constraints \eqref{eq3_PF_1} and \eqref{eq3_PF_2} represent the active and reactive power limits of the lines, where the limits ($P_{k}^{\rm K,M}$ and $Q_{k}^{\rm K,M}$) are multiplied by the line status binary variable $x_{k,t}^{\rm K}$. Therefore, if a line is disconnected or damaged $x_{k,t}^{\rm K}=0$, then constraints \eqref{eq3_PF_1} and \eqref{eq3_PF_2} will be relaxed, which means that power cannot flow through this line. In the proposed model, there are two types of IBDGs, grid-forming IBDGs with black start capability and grid-following IBDGs without black start capability. On the one side, the grid-forming IBDGs can provide voltage and frequency references in the MG during the restoration process, which can energize the bus and restore the part of the network that is not damaged if the fault is isolated. Therefore, the grid-forming IBDGs are considered to be connected to the network at the beginning of restoration. On the other side, the grid-following IBDGs are switched off at the beginning of restoration. If the grid-following IBDGs are connected to an energized bus during the restoration process, then they can be switched on to supply active and reactive powers. In constraints \eqref{eq3_DG_1} and \eqref{eq3_DG_2}, the active and reactive power outputs of the grid-forming IBDGs are limited by the maximum active and reactive capacities $P_{i}^{\rm G,M}$ and $Q_{i}^{\rm G,M}$, respectively. Constraints \eqref{eq3_DG_3} and \eqref{eq3_DG_4} limit the active and reactive outputs of the grid-following IBDGs. Note that the constraints \eqref{eq3_DG_3} and \eqref{eq3_DG_4} of grid-following IBDGs are multiplied by binary variable $x_{i,t}^{\rm G}$. Consequently, if one grid-following IBDG is not energized ($x_{i,t}^{\rm G}=0$) during the restoration process, then constraints \eqref{eq3_DG_3} and \eqref{eq3_DG_4} of this grid-following IBDG will be relaxed.
\begin{align}\label{eq3_bal_1}
\sum_{k\in \Omega_{\rm K}(i,.)}P_{k,\phi,t}^{\rm K} - \sum_{k\in \Omega_{\rm K}(.,i)}P_{k,\phi,t}^{\rm K} = P^{\rm G}_{i,\phi,t}-x_{i,t}^{\rm L} P_{i,\phi,t}^{\rm L},\forall i,\phi,t
\end{align}  
\begin{align}\label{eq3_bal_2}
\sum_{k\in \Omega_{\rm K}(i,.)}Q_{k,\phi,t}^{\rm K} - \sum_{k\in \Omega_{\rm K}(.,i)}Q_{k,\phi,t}^{\rm K} &= Q^{\rm G}_{i,\phi,t} - x_{i,t}^{\rm L} Q_{i,\phi,t}^{\rm L},\forall i,\phi,t
\end{align}
\begin{equation}\label{eq3_PF_1}
- x_{k,t}^{\rm K} P_{k}^{\rm K,M}\leq P_{k,\phi,t}^{\rm K}\leq x_{k,t}^{\rm K} P_{k}^{\rm K,M},\forall k \in \Omega_{\rm K},\phi,t
\end{equation}
\begin{equation}\label{eq3_PF_2}
-x_{k,t}^{\rm K} Q_{k}^{\rm K,M}\leq Q_{k,\phi,t}^{\rm K}\leq x_{k,t}^{\rm K} Q_{k}^{\rm K,M},\forall k \in \Omega_{\rm K},\phi,t
\end{equation}
\begin{equation}\label{eq3_DG_1}
0 \leq P_{i,\phi,t}^{\rm G} \leq P_{i}^{\rm G,M} ,\forall i \in \Omega_{\rm BS},\phi,t
\end{equation}
\begin{equation}\label{eq3_DG_2}
0\leq Q_{i,\phi,t}^{\rm G} \leq Q_{i}^{\rm G,M} ,\forall i \in \Omega_{\rm BS},\phi,t
\end{equation}
\begin{equation}\label{eq3_DG_3}
0 \leq P_{i,\phi,t}^{\rm G} \leq x_{i,t}^{\rm G} P_{i}^{\rm G,M} ,\forall i \in \Omega_{\rm NBS},\phi,t
\end{equation}
\begin{equation}\label{eq3_DG_4}
0\leq Q_{i,\phi,t}^{\rm G} \leq x_{i,t}^{\rm G} Q_{i}^{\rm G,M} ,\forall i \in \Omega_{\rm NBS},\phi,t
\end{equation}

Constraints \eqref{eq3_U_1} and \eqref{eq3_U_2} calculate the voltage difference along line $k$ between bus $i$ and bus $j$, where $U_{i,\phi,t}$ is the square of voltage magnitude of bus $i$. We use the big-M method \cite{Restoration_6} to relax constraints \eqref{eq3_U_1} and \eqref{eq3_U_2}, if lines are damaged or disconnected, then $x_{k,t}^{\rm K}=0$. The $p_{k,\phi}$ represents the phase identifier for phase $\phi$ of line $k$. For example, if line $k$ is a single-phase line on phase a, then $p_{k,\phi_a}=1$, $p_{k,\phi_b}=0$ and $p_{k,\phi_c}=0$. Constraint \eqref{eq3_U_3} guarantees that the voltage is limited within a specified region [$U^{\rm m}_{i}$,$U^{\rm M}_{i}$], and will be set to 0 if the bus is in an outage area $x_{i,t}^{\rm B}=0$. 
\begin{equation}\label{eq3_U_1}
\begin{split}
 U_{i,\phi,t}-U_{j,\phi,t} \geq &2(\hat{R}_{k}P_{k,\phi,t}^{\rm K}+\hat{X}_{k}Q_{k,\phi,t}^{\rm K})\\
&+ (x_{k,t}^{\rm K}+p_{k,\phi}-2)M,\forall k,ij \in \Omega_{\rm K},\phi,t
\end{split}
\end{equation}
\begin{equation}\label{eq3_U_2}
\begin{split}
U_{i,\phi,t}-U_{j,\phi,t} \leq &2(\hat{R}_{k}P_{k,\phi,t}^{\rm K}+\hat{X}_{k}Q_{k,\phi,t}^{\rm K})\\
&+ (2-x_{k,t}^{\rm K}-p_{k,\phi})M,\forall k,ij \in \Omega_{\rm K},\phi,t
\end{split}
\end{equation}
\begin{equation}\label{eq3_U_3}
x_{i,t}^{\rm B} U^{\rm m}_{i} \leq U_{i,\phi,t} \leq x_{i,t}^{\rm B} U^{\rm M}_{i},\forall i,\phi,t
\end{equation}
where $\hat{R}_{k}$ and $\hat{X}_{k}$ are the unbalanced three-phase resistance matrix and reactance matrix of line $k$. To model the unbalanced three-phase network, we assume that the distribution network is not too severely unbalanced and operates around the nominal voltage, then the relative phase unbalance can be approximated as $a_{\phi}=[1, {e}^{-{\bf i}2\pi/3},  {e}^{{\bf i}2\pi/3}]^T$ \cite{unbal}. Therefore, the equivalent unbalanced three-phase system line impedance matrix $\hat{Z}_{k}$ can be calculated based on the original line impedance matrix $Z_{k}$ and $a_{\phi}$ in \eqref{unbal_Z}. $\hat{R}_{k}$ and $\hat{X}_{k}$ are the real and imaginary parts of $\hat{Z}_{k}$, as shown in \eqref{unbal_RX}. Note that the loads and IBDGs are also modelled in a three-phase form. More details about the model of unbalance three-phase distribution system can be found in \cite{QZ_CVR}.    
\begin{equation}
	\hat{Z}_{k}=a_{\phi}a_{\phi}^{H}\odot Z_{k}\label{unbal_Z}
\end{equation}
\begin{equation}
	\hat{R}_{k}=real(\hat{Z}_{k}),\hat{X}_{k}=imag(\hat{Z}_{k})\label{unbal_RX}
\end{equation}
     
Constraints \eqref{eq3_DG_5}-\eqref{eq3_nottrip_load} ensure the physical connections among buses, lines, IBDGs and loads during restoration process. In constraint \eqref{eq3_DG_5}, the grid-following IBDGs will be switched on $x_{i,t}^{\rm G}=1$, if the connected bus is energized $x_{i,t}^{\rm B}=1$; otherwise, $x_{i,t}^{\rm G}=0$. Constraint \eqref{eq3_br_1} implies a switchable line can only be energized when both end buses are energized. Constraint \eqref{eq3_br_3} presents that a non-switchable line can be energized once one of two end buses is energized. Constraint \eqref{eq3_load_1} ensures that a switchable load can be energized $x_{i,t}^{\rm L}=1$, if the connected bus is energized $x_{i,t}^{\rm B}=1$; otherwise, $x_{i,t}^{\rm L}=0$. Constraint \eqref{eq3_load_2} allows that a non-switchable load can be immediately energized once the connected bus is energized. Constraints \eqref{eq3_nottrip_DG}-\eqref{eq3_nottrip_load} ensure that the grid-following IBDGs, switchable lines and loads cannot be tripped again, if they have been energized at the previous time $t-1$. 
\begin{equation}
x_{i,t}^{\rm G} \leq x_{i,t}^{\rm B} ,\forall i\in \Omega_{\rm NBS},t \label{eq3_DG_5}
\end{equation}
\begin{equation}
x_{k,t}^{\rm K} \leq x_{i,t}^{\rm B}, x_{k,t}^{\rm K} \leq x_{j,t}^{\rm B}, \forall k, ij\in \Omega_{\rm SW_K},t\label{eq3_br_1}
\end{equation}
\begin{equation}
x_{k,t}^{\rm K} = x_{i,t}^{\rm B}, x_{k,t}^{\rm K} = x_{j,t}^{\rm B} ,\forall k, ij\in \Omega_{\rm NSW_K},t\label{eq3_br_3}
\end{equation}
\begin{equation}
x_{i,t}^{\rm L} \leq x_{i,t}^{\rm B} ,\forall i\in \Omega_{\rm SW_L},t \label{eq3_load_1}
\end{equation}
\begin{equation}
x_{i,t}^{\rm L} = x_{i,t}^{\rm B}, \forall i\in \Omega_{\rm NSW_L},t\label{eq3_load_2}
\end{equation}
\begin{equation}
x_{i,t}^{\rm G} - x_{i,t-1}^{\rm G} \geq 0 ,\forall i\in \Omega_{\rm NBS},t \label{eq3_nottrip_DG}
\end{equation}
\begin{equation}
x_{k,t}^{\rm K} - x_{k,t-1}^{\rm K} \geq 0 ,\forall k\in \Omega_{\rm SW_k},t \label{eq3_nottrip_line}
\end{equation}
\begin{equation}
x_{i,t}^{\rm L} - x_{i,t-1}^{\rm L} \geq 0 ,\forall i\in \Omega_{\rm SW_L},t \label{eq3_nottrip_load}
\end{equation}

Constraints \eqref{eq3_radial_1}-\eqref{eq3_radial_3} ensure that each formed MG remains isolated from each other and each MG can maintain a tree topology during the restoration process. Constraint \eqref{eq3_radial_1} implies that if one bus $i$ is located in one bus block, $i\in \Omega_{\rm BK}$, then the energization status of bus and the corresponding bus block keep the same. Here $x_{B,t}^{\rm BK}$ represents the energization status of bus block $BK$. To avoid forming loop topology, constraint \eqref{eq3_radial_2} guarantees that a switchable line cannot be closed at time $t$ if its both end bus blocks are already energized at previous time $t-1$. Note that the DistFlow model is valid for radial distribution network, therefore, loop topology is not considered in this work. If one bus block is not energized at previous time $t-1$, then constraint \eqref{eq3_radial_3} makes sure that this bus block can only be energized at time $t$ by at most one of the connected switchable lines. Constraints \eqref{eq3_seq_1} and \eqref{eq3_seq_2} ensure that each formed MG has a reasonable restoration and energization sequence of switchable lines and bus blocks. Constraints \eqref{eq3_seq_1} implies that energized switchable lines can energize the connected bus block. Constraints \eqref{eq3_seq_2} requires that a switchable line can only be energized at time $t$, if at least one of the connected bus block is energized at previous time $t-1$. 
\begin{equation}
x_{i,t}^{\rm B} = x_{i,t}^{\rm BK}, \forall i\in \Omega_{\rm BK},t \label{eq3_radial_1}
\end{equation}
\begin{equation}\label{eq3_radial_2}
\begin{split}
(x_{i,t}^{\rm BK}-x_{i,t-1}^{\rm BK}) &+ (x_{j,t}^{\rm BK}-x_{j,t-1}^{\rm BK})\\
&\geq x^{\rm K}_{k,t} - x^{\rm K}_{k,t-1}, \forall k, ij\in \Omega_{\rm SW_K},t \geq 2
\end{split}
\end{equation}
\begin{equation}\label{eq3_radial_3}
\begin{split}
\sum_{ki, k\in \Omega_{i}}(x^{\rm K}_{ki,t}-x^{\rm K}_{ki,t-1})&+\sum_{ij, j\in \Omega_{i}}(x^{\rm K}_{ij,t}-x^{\rm K}_{ij,t-1})\\
&\leq 1+x_{i,t-1}^{\rm BK}M, \forall k, ij\in \Omega_{\rm SW_K},t \geq 2
\end{split}
\end{equation}
\begin{equation}
x_{i,t-1}^{\rm BK} \leq \sum_{ki, k\in \Omega_{i}}(x^{\rm K}_{ki,t})+\sum_{ij, j\in \Omega_{i}}(x^{\rm K}_{ij,t}), \forall k, ij\in \Omega_{\rm SW_K},t \geq 2 \label{eq3_seq_1}
\end{equation}
\begin{equation}
x^{\rm K}_{ij,t}\leq x_{i,t-1}^{\rm BK}+x_{j,t-1}^{\rm BK}, \forall ij\in \Omega_{\rm SW_K},t \geq 2 \label{eq3_seq_2}
\end{equation}

\subsection{Simulation-based Frequency Dynamics Constraints}\label{sec:MLS}
By considering the frequency dynamics of each isolated inverter-dominated MG during the transitions of network reconfiguration and service restoration, constraints \eqref{eq3_MSL} and \eqref{eq3_DG_ramp} have been added here to avoid the potential large frequency deviations caused by MG formation and oversized load restoration. The variable of maximum load step $P_{i,t}^{G,MLS}$ has been applied in constraint \eqref{eq3_MSL} to ensure that the restored load is limited by an upper bound for each restoration stage, as follows:   
\begin{equation}\label{eq3_MSL}
\begin{split}
 0\leq P_{i,t}^{\rm G,MLS} &\leq P_{i,t-1}^{\rm G,MLS}\\
&+ \alpha (\Delta f^{\rm max} - \Delta f^{\rm meas}),\forall i\in \Omega_{\rm BS},t \geq 2
\end{split}
\end{equation}

In constraint \eqref{eq3_MSL}, the variable $P_{i,t}^{G,MLS}$ is restricted by three items: a hyper-parameter $\alpha$ representing the virtual frequency-power characteristic of IBDGs, a user-defined maximum allowable frequency drop limit $\Delta f^{\rm max}$ and the measured maximum transient frequency drop from the results of simulation level $\Delta f^{\rm meas}$. The hyper-parameter $\alpha$ is used to curb the frequency nadir during transients from too low. This can be shown by the following expressions:
\begin{align}\label{eq3_alpha}
\nonumber \alpha (\Delta f^{\rm max} - \Delta f^{\rm meas})&= \alpha (f_0 - f^{\rm min} - (f_0 - f^{\rm nadir}))\\
\nonumber &=\alpha (f^{\rm nadir} - f^{\rm min})\\
&\triangleq \Delta P_{i,t-1}^{\rm G,MLS}
\end{align}
where $f_0$ is the nominal steady-state frequency, e.g. 60Hz. $f^{\rm nadir}$ is the lowest frequency reached during the transient simulation. $f^{\rm min}$ is the minimum allowable frequency. $\Delta P_{i,t-1}^{G,MLS}$ is the incremental change of the maximum load step for the next step $t$ (estimated at step $t-1$). Finally, constraint \eqref{eq3_DG_ramp} ensures the restored load and frequency response of the IBDGs do not exceed the user-defined thresholds.  
\begin{equation}\label{eq3_DG_ramp}
\begin{split}
-x_{i,t}^{\rm G} P_{i,t}^{\rm G,MLS} \leq P_{i,\phi,t}^{\rm G}-P_{i,\phi,t-1}^{\rm G} \leq & x_{i,t}^{\rm G} P_{i,t}^{\rm G,MLS}\\
&, i\in \Omega_{\rm BS},\phi,t \geq 2
\end{split}
\end{equation}

Note that the generator ramp rate is not a constant number anymore as in previous literature, but is varying with the value of $P_{i,t}^{\rm G,MLS}$ from \eqref{eq3_MSL} during the optimization process combining with transient simulation information of frequency deviation. When $f^{\rm nadir}$ is approaching $f^{\rm min}$, that implies a necessity to reduce the potential amount of restored load in the next step. Thus the incremental change of maximum load step $\Delta P_{i,t}^{\rm G,MLS}$ is reduced to reflect the above purpose. During the restoration process, the restored load in each restoration stage is determined by maximum load step and available DG power output through power balance constraints \eqref{eq3_bal_1}, \eqref{eq3_bal_2} and constraints \eqref{eq3_MSL}, \eqref{eq3_DG_ramp} in optimization level; then, the frequency deviation in each restoration stage is determined by restored load through transient model in simulation level, which is introduced in the next section. 

\section{Transient Simulation of Inverter-Dominated MG Formation}\label{sec:Level_II}
In optimization level, our target is to maximize the amount of restored load while satisfying a series of constraints. One of these constraints should be frequency dynamics constraint which is derived from simulation level. However, due to the different time scales and nonlinearity, the conventional dynamic security constraints cannot be directly solved in optimization problem, such as Lyapunov theory, LaSalle's theorem and so on. Therefore, we need a connection variable between the two levels. 

For this purpose, we assume that the changes of typologies between each two sequential stages can be represented by the change of restored loads $P^{\rm L}$. The sudden load change of $P^{\rm L}$ results in a disturbance in MGs in the time-scale of simulation level. During the transience to the new equilibrium (operation point), the system states such as frequency will deviate from their nominal values. Therefore, it is natural to estimate the dynamic security margin with the allowed maximum range of deviations. 
\begin{figure} 
	\vspace{-0pt} 
	\vspace{-0pt}
	\centering
	\includegraphics[width=1.0\linewidth]{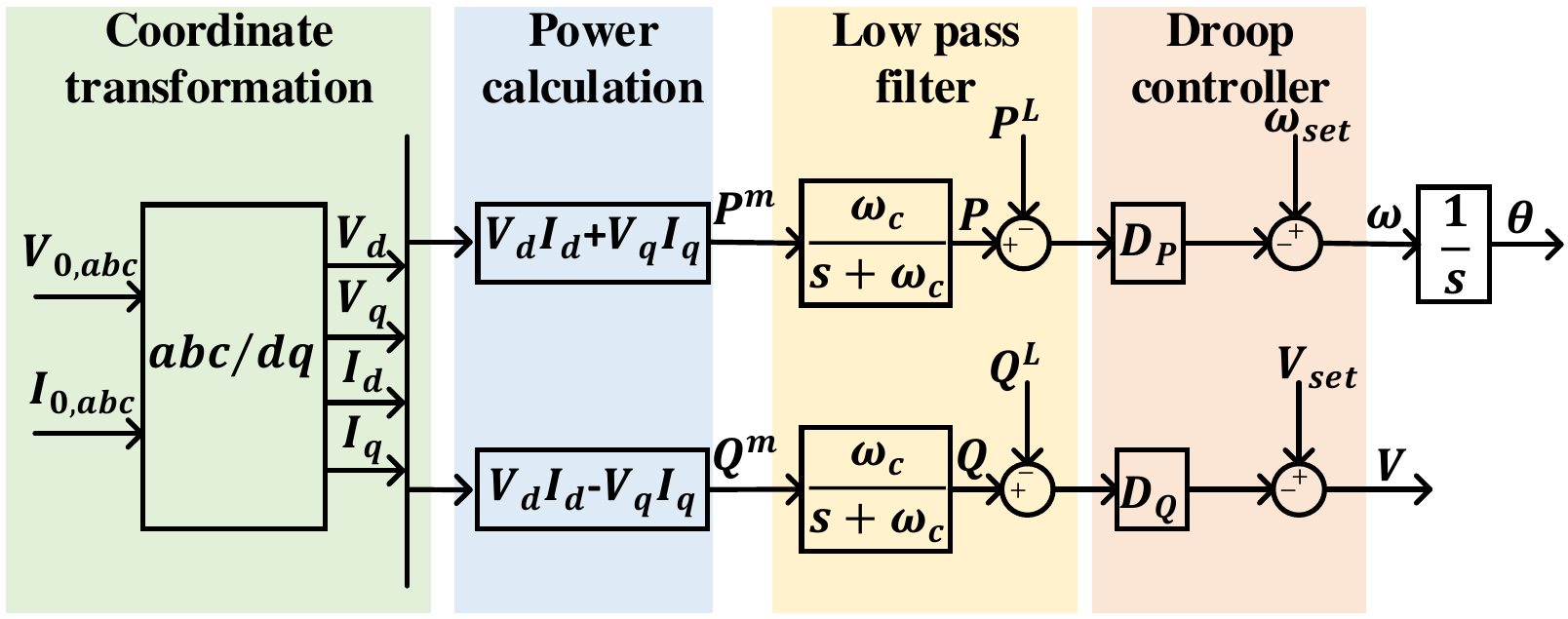}
	\vspace{-0pt}\
	\caption{Diagram of studied MG control system.}  
	\centering
	\label{droop}
		\vspace{-0pt}\
\end{figure}

Since the frequency of each inverter-dominated MG is mainly controlled by the grid-forming IBDGs, we can approximate the maximum frequency deviation during the transience by observing the dynamic response of the grid-forming IBDGs under sudden load change. In this paper, the standard outer droop control together with inner double-loop control structure is adopted for each IBDGs unit. 
As shown in Fig. \ref{droop}, the three-phase output voltage $V_{0,abc}$ and current $I_{0,abc}$ are measured from the terminal bus of the inverter and transformed into $dq$ axis firstly. Then, the filtered terminal output active and reactive power $P$ and $Q$ are obtained by filtering the calculated power measurements $P^{\rm meas}$ and $Q^{\rm meas}$ with cut-off frequency $\omega_{\rm c}$. Finally, the voltage and frequency references for the inner control loop are calculated with droop controller. Since the references can be accurately tracked by inner control loop with properly tuned PID parameters in the much faster time-scale, the output voltage $V$ and frequency $\omega$ can be considered equivalently as the references generated by the droop controller. Thus, the inverter can be modelled effectively modelled by using the terminal states and line states of the inverter \cite{Inverters_model_1,Inverters_model_3}. In this work, the transient simulation is conducted with the detailed mathematical MG model (31)--(37) adopted from \cite{Inverters_model_1}, where the droop equations (34) and (35) are replaced by the ones proposed in \cite{Inverters_model_3} to consider the restored loads.
\begin{align}
    \dot{P}&=\omega_{\rm c} (V\cos{\theta}I_{\rm d}+V\sin{\theta}I_{\rm q}-P),\\
    \dot{Q}&=\omega_{\rm c} (V\sin{\theta}I_{\rm d}-V\cos{\theta}I_{\rm q}-Q),\\
    \dot{\theta}&=\omega-\omega_0,\\
    \dot{\omega}&=\omega_{\rm c} (\omega_{\rm set}-\omega+D_{\rm P}(P-P^{\rm L})),\\
    \dot{V}&=\omega_{\rm c} (V_{\rm set}-V+D_{\rm Q}(Q-Q^{\rm L})),\\
    \dot{I}_{\rm d}&=(V\cos{\theta}-V_{\rm bus}-R I_{\rm d})/L+\omega_o I_{\rm q},\\
    \dot{I}_{\rm q}&=(V\sin{\theta}-R I_{\rm q})/L-\omega_o I_{\rm d},
\end{align}
where $\omega_{\rm set}$ and $V_{\rm set}$ are the set points of frequency and voltage controllers, respectively; $\omega_{\rm c}$ is cut-off frequency; $D_{\rm P}$ and $D_{\rm Q}$ are $P-\omega$ and $Q-V$ droop gains, respectively; $P^{\rm L}$ and $Q^{\rm L}$ are the restored active and reactive loads, respectively; $\theta$ is phase angle; $\omega$ is angular frequency in $rad/s$; $\omega_0$ is a fixed angular frequency; $V_{\rm bus}$ is bus voltage; $I_{\rm d}$ and $I_{\rm q}$ are $dq$-axis currents; $R$ and $L$ are aggregate resistance and inductance of connections from the inverter terminal's point view, respectively. In (34), it can be observed that, the equilibrium can be achieved when $\omega=\omega_{\rm set}$ and $P=P^{\rm L}$, which means that the output frequency tracks the frequency reference when the output power of the simulation level tracks the obtained restored load of the optimization level. 

Note that constraint \eqref{eq3_MSL} is the connection between the optimization level and simulation level in our proposed two-level simulation-assisted restoration model, which incorporates the frequency response of inverter-dominated MG from the simulation level into the optimization level. The variable $P_{i,t}^{\rm G,MLS}$ is restricted by frequency response in constraint \eqref{eq3_MSL}. Meanwhile, $P_{i,t}^{\rm G,MLS}$ also limits the IBDG power output in constraint \eqref{eq3_DG_ramp}. In constraints \eqref{eq3_bal_1} and \eqref{eq3_bal_2}, the power balance is met between restored load and power supply of IBDGs. Therefore, we associate the frequency nadir of the transient simulation with respect to the restored load by incorporating the frequency dynamics constraints explicitly in the optimization level.

After the process of fault detection \cite{Fault_dec} and sub-grids isolation are finished, the proposed service restoration model will begin to work. Each isolated network will begin to form a MG depending on the location of the nearest grid-forming IBDG with black start capability. The flowchart of the proposed restoration method is shown in Fig. \ref{flowchart} and the interaction between the proposed transient simulation and the established optimization problem of service restoration is described as follows:
\begin{figure}
	\vspace{-0pt} 
	\vspace{-0pt}
	\centering
	\includegraphics[width=1.0\linewidth]{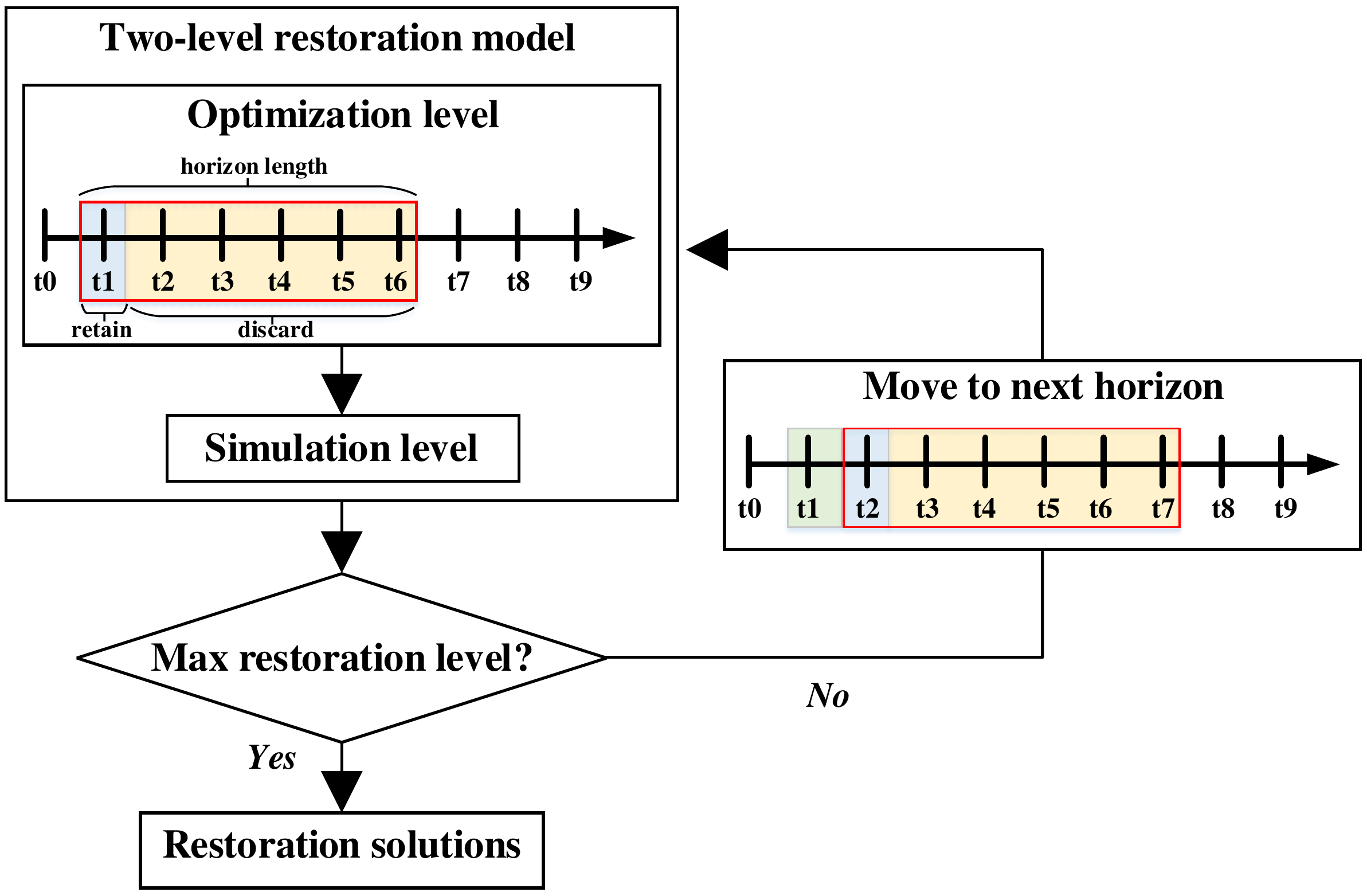}
	\vspace{-0pt} 
	\caption{Flowchart of the proposed two-level simulation-assisted restoration method.}
	\centering
	\label{flowchart}
    \vspace{-0pt} 
\end{figure} 

\textbf{(a)} \textit{Solving the optimal service restoration problem}: Given horizon length $T$ in each restoration stage, the MILP-based sequential service restoration problem \eqref{eq3_obj_1}--\eqref{eq3_MSL} and \eqref{eq3_DG_ramp} is solved, and the restoration solution is obtained for each formed MG. 

\textbf{(b)} \textit{Transient simulation of inverter-dominated MGs}: After receiving restoration solutions of current stage from optimization level, the frequency response is simulated by (31)--(37) and the frequency nadir is calculated for each inverter-dominated MG.

\textbf{(c)} \textit{Check the progress of service restoration and stopping criteria}: If the maximum service restoration level is reached for all the MGs, then stop the restoration process; otherwise, go back to (a) to generate the restoration solution with newly obtained frequency responses of all MGs for next restoration stage. 

\section{Numerical Results}\label{sec:Results}
\subsection{Simulation Setup}
\begin{figure} 
	\vspace{-0pt} 
	\vspace{-0pt}
	\centering
	\includegraphics[width=1.0\linewidth]{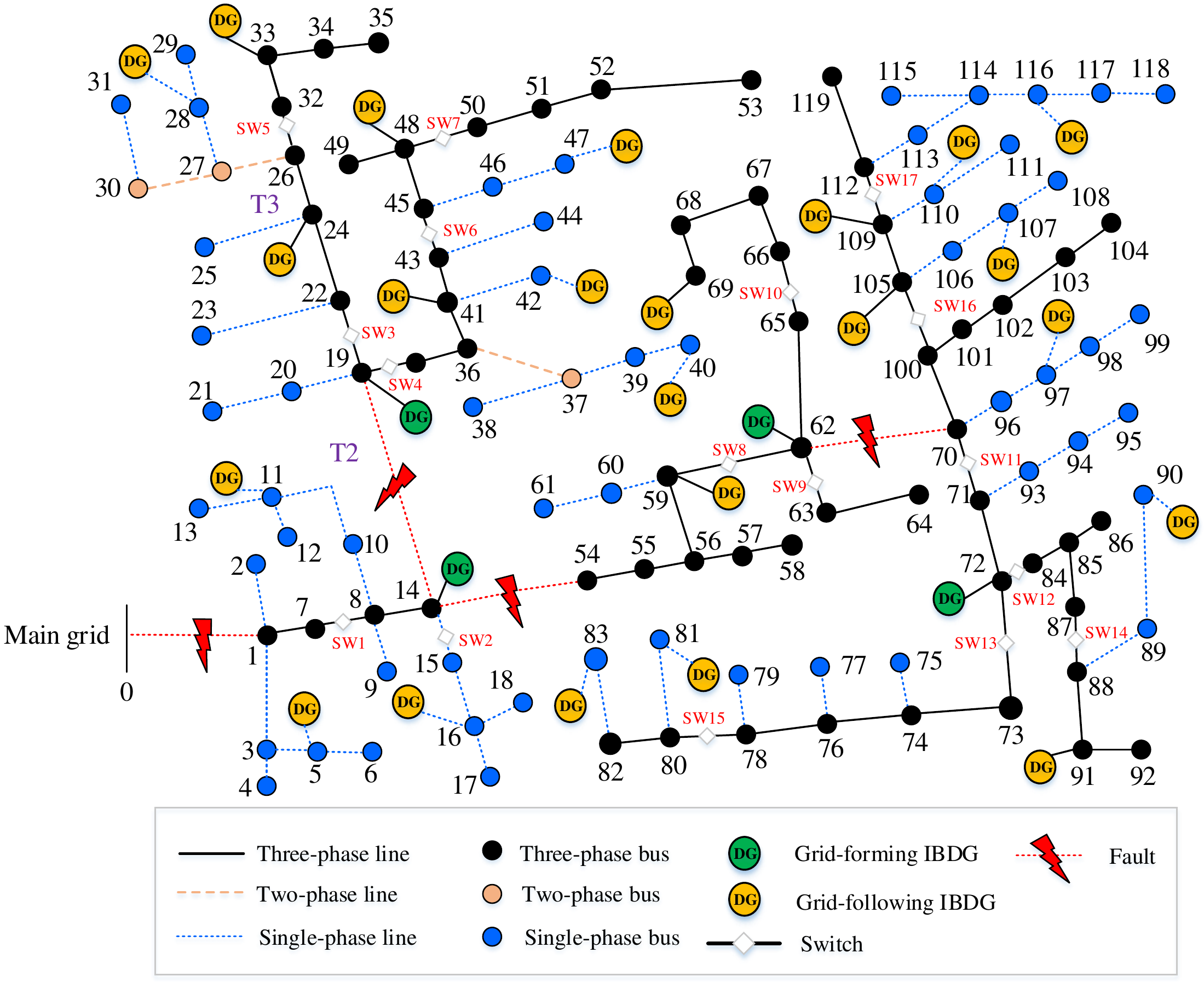}
	\vspace{-0pt}\
	\caption{Modified IEEE 123 node test feeder.} 
	\centering
	\label{test_system_1}
		\vspace{-0pt}\
\end{figure}

A modified IEEE 123-bus test system \cite{123bus_system} in Fig. \ref{test_system_1} is used to test the performance of the proposed frequency dynamics constrained service restoration model. In Fig. \ref{test_system_1}, blue dotted line and blue dot stand for single-phase line and bus, orange dashed line and orange dot stand for two-phase line and bus, black line and black dot stand for three-phase line and bus, respectively. The modified test system has been equipped with multiple remotely controlled switches, as shown in Fig. \ref{test_system_1}. In Table \ref{123_bus_devices}, the locations and capacities of grid-following and grid-forming IBDGs are shown. Four line faults on lines between substation and bus 1, bus 14 and bus 19, bus 14 and bus 54 and bus 62 and bus 70 are detected, as shown in red dotted lines of Fig. \ref{test_system_1}. They are assumed to be persisting during the restoration process until the faulty areas are cleared to maintain the radial topology and isolate the faulty areas. Consequently, four MGs can be formed for service restoration with grid-forming IBDGs and switches. For the sake of simplicity, we assume that the weight factors for all loads are set to 1 during the restoration process. We demonstrate the effectiveness of our proposed service restoration model through numerical evaluations on the following experiments: (i) Comparison between a base case (i.e., without the proposed frequency dynamics constraints) and the case with the proposed restoration model. (ii) Cases with the proposed restoration model under different values of hyper-parameters. All the case studies are implemented using a PC with Intel Core i7-4790 3.6 GHz CPU and 16 GB RAM hardware. The simulations are performed in MATLAB R2019b, which integrates YALMIP Toolbox with IBM ILOG CPLEX 12.9 solver and ordinary differential equation solver.   

\begin{table}[]
    \centering
	\renewcommand{\arraystretch}{1.3}	
	\caption{Locations and Capacities of Grid-following and Grid-forming IBDGs in modified IEEE 123 Node Test Feeder.}
	\label{123_bus_devices}
	\begin{tabular}{lll}
		\hline\hline
		Type                     & Locations                                                                                                                     & Capacities                                                                                                    \\ \hline
		\multicolumn{1}{l}{\begin{tabular}[l]{@{}l@{}}Grid-following \\IBDG (1-$\phi$)\end{tabular}}                       & \begin{tabular}[l]{@{}l@{}}5, 11, 16, 28, 40, 42, \\ 47, 81, 83, 90, 97, 107\\ 110, 116\end{tabular} & \begin{tabular}[l]{@{}l@{}}80 kW for single-phase \\ 40 kVAr for single-phase \end{tabular}                                                                       \\ \hline
	    \multicolumn{1}{l}{\begin{tabular}[l]{@{}l@{}}Grid-following \\IBDG (3-$\phi$)\end{tabular}}                       & \begin{tabular}[l]{@{}l@{}}24, 33, 41, 48, 52,\\ 59, 69, 91, 105, 109\end{tabular}                                                                                                                   & \begin{tabular}[l]{@{}l@{}}100 kW per $\phi_{a}, \phi_{b}, \phi_{c}$ \\ 50 kVAr per $\phi_{a}, \phi_{b}, \phi_{c}$ \end{tabular} \\\hline
	    \multicolumn{1}{l}{\begin{tabular}[l]{@{}l@{}}Grid-forming \\IBDG (3-$\phi$)\end{tabular}}                       & 14, 19, 62, 72                                                                                                                   & \begin{tabular}[l]{@{}l@{}}100 kW per $\phi_{a}, \phi_{b}, \phi_{c}$ \\ 50 kVAr per $\phi_{a}, \phi_{b}, \phi_{c}$ \end{tabular} \\
	    \hline\hline
	\end{tabular}
\end{table}

\subsection{Sequential Service Restoration Results}
As shown in \eqref{eq3_MSL}, the relationship between the maximum load step and the frequency nadir is influenced by the value of hyper-parameter $\alpha$ in the frequency-dynamics constraints. Therefore, different $\alpha$ values may lead to different service restoration results. In this case, the horizon length $T$ and the hyper-parameter $\alpha$ are set to 4 and 0.1, respectively.

\begin{figure} 
	\vspace{-0pt} 
	\vspace{-0pt}
	\centering
	\includegraphics[width=1.0\linewidth]{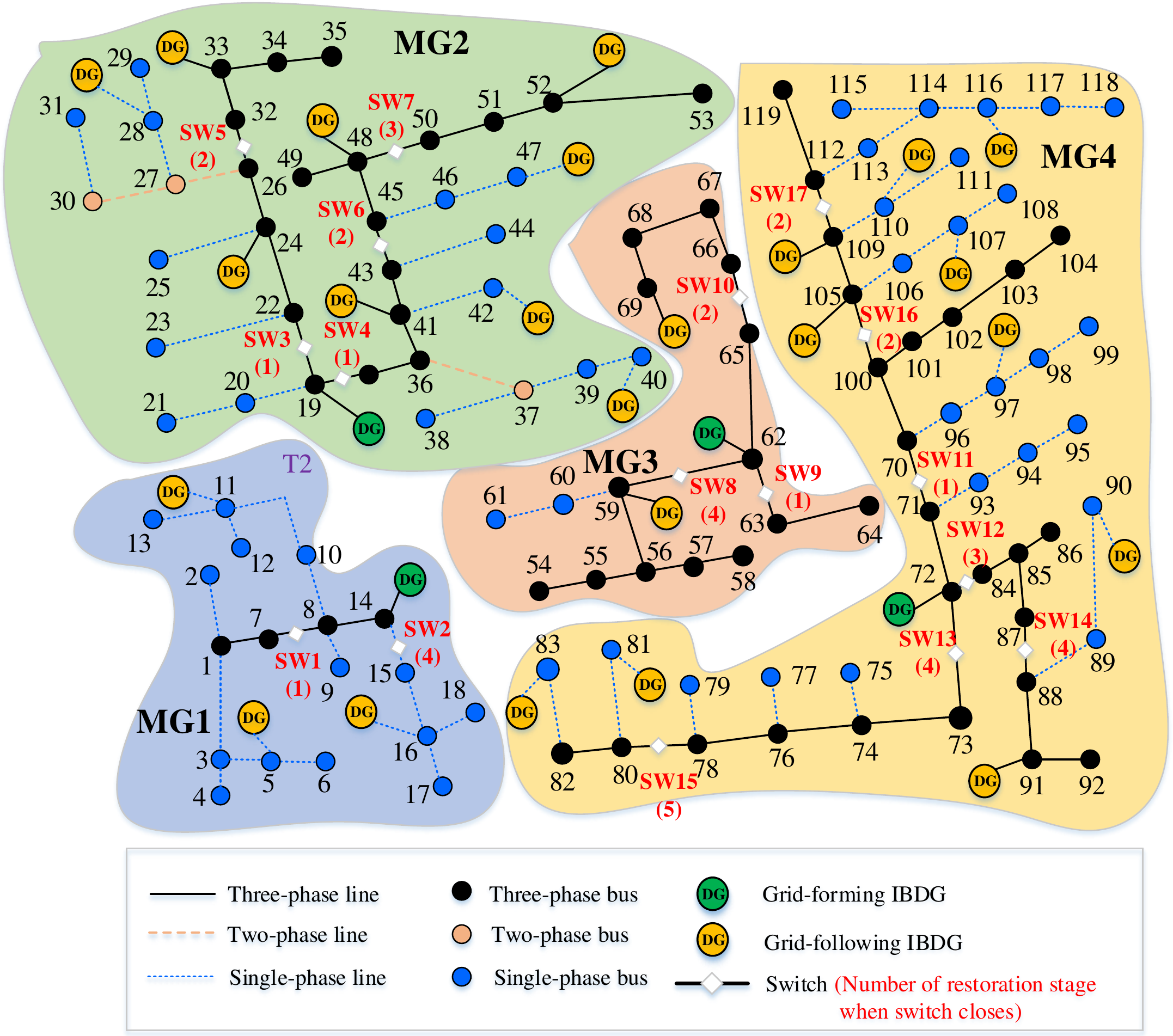}
	\vspace{-0pt}\
	\caption{Restoration solutions for the formed MG1-MG4, where the restoration stage when line switch closes is shown in red.}  
	\centering
	\label{results_123}
		\vspace{-0pt}\
\end{figure}

As shown in Fig. \ref{results_123}, the system is partitioned into four MGs by energizing the switchable lines sequentially, and the radial structure of each MG is maintained at each stage. Inside each formed MG, the power balance is achieved between the restored load and power outputs of IBDGs. The value in brackets nearby each line switch in Fig. \ref{results_123} represents the number of restoration stage when it closes. In Table \ref{results_123_table}, the restoration sequences for switchable IBDGs and loads are shown, where the subscript and superscript are the bus index and the MG index of grid-following IBDGs and loads, respectively. It can be observed that MG2 only needs 3 stages to be fully restored, while MG1 and MG3 can restore in 4 stages. However, due to the heavy loading situation, MG4 is gradually restored in 5 stages to ensure a relatively smooth frequency dynamics. 

\begin{table}[]
    \centering
	\renewcommand{\arraystretch}{1.3}	
	\caption{Restored Grid-following IBDGs and Loads at Each Restoration Stage.}
	\label{results_123_table}
	\begin{tabular}{lll}
		\hline\hline
		\begin{tabular}[l]{@{}l@{}}Restoration \\ stage\end{tabular}                     & \begin{tabular}[l]{@{}l@{}}Restored \\ grid-following IBDGs\end{tabular}                                                                                                                      & \begin{tabular}[l]{@{}l@{}}Restored \\ loads\end{tabular}                                                                                                    \\ \hline
		1                       & \begin{tabular}[l]{@{}l@{}}$G_{11}^1,G_5^1,G_{24}^2$ \\ $G_{28}^2,G_{40}^2,G_{41}^2$\\ $G_{42}^2,G_{97}^4$\end{tabular} & \begin{tabular}[l]{@{}l@{}}$L_{14}^1,L_8^1,L_9^1,L_{10}^1,L_{11}^1,L_{12}^1$\\ $L_{13}^1,L_1^1,L_2^1,L_3^1,L_4^1,L_5^1,L_6^1$ \\ $L_7^1,L_{19}^2,L_{20}^2,L_{21}^2,L_{22}^2,L_{23}^2$\\ $L_{24}^2,L_{25}^2,L_{26}^2,L_{27}^2L_{28}^2,L_{29}^2$\\ $L_{30}^2,L_{31}^2,L_{36}^2,L_{37}^2L_{38}^2,L_{39}^2$\\ $L_{40}^2,L_{41}^2,L_{42}^2,L_{43}^2,L_{44}^2,L_{62}^3$\\ $L_{65}^3,L_{63}^3,L_{64}^3,L_{72}^4,L_{71}^4,L_{93}^4$\\ $L_{94}^4,L_{95}^4,L_{70}^4,L_{96}^4,L_{97}^4,L_{98}^4$\\$L_{99}^4,L_{100}^4,L_{101}^4,L_{102}^4,L_{103}^4$\\$L_{104}^4$ \end{tabular} \\
		\hline
	    2                       & \begin{tabular}[l]{@{}l@{}}$G_{33}^2,G_{47}^2,G_{48}^2$ \\ $G_{69}^3,G_{105}^4,G_{107}^4$\\ $G_{109}^4,G_{110}^4,G_{116}^4$\end{tabular}                                                                                                                  & \begin{tabular}[l]{@{}l@{}}$L_{32}^2,L_{33}^2,L_{34}^2,L_{35}^2,L_{45}^2,L_{46}^2$ \\ $L_{47}^2,L_{48}^2,L_{49}^2,L_{66}^3,L_{67}^3,L_{68}^3$\\ $L_{69}^3,L_{105}^4,L_{106}^4,L_{107}^4,L_{108}^4$\\ $L_{109}^4,L_{110}^4,L_{111}^4,L_{112}^4,L_{113}^4$\\ $L_{114}^4,L_{115}^4,L_{116}^4,L_{117}^4,L_{118}^4$ \\ $L_{119}^4$\end{tabular} \\\hline
	    3                       & \begin{tabular}[l]{@{}l@{}}$G_{52}^2$\end{tabular}                                                                                                                   & \begin{tabular}[l]{@{}l@{}}$L_{50}^2,L_{51}^2,L_{52}^2,L_{53}^2,L_{84}^4,L_{85}^4$\\ $L_{86}^4,L_{87}^4$ \end{tabular} \\\hline
	    4                       & \begin{tabular}[l]{@{}l@{}}$G_{16}^1,G_{59}^3,G_{90}^4$ \\ $G_{91}^4$\end{tabular}                                                                                                                   & \begin{tabular}[l]{@{}l@{}} $L_{15}^1,L_{16}^1,L_{17}^1,L_{18}^1,L_{54}^3,L_{55}^3$\\ $L_{56}^3,L_{57}^3,L_{58}^3,L_{59}^3,L_{60}^3,L_{60}^3$\\ $L_{61}^3,L_{73}^4,L_{74}^4,L_{75}^4,L_{76}^4,L_{77}^4$\\ $L_{78}^4,L_{79}^4,L_{88}^4,L_{89}^4,L_{90}^4,L_{91}^4$\\ $L_{92}^4$ \end{tabular} \\\hline
	    5                       & \begin{tabular}[l]{@{}l@{}}$G_{81}^4,G_{83}^4$\end{tabular}                                                                                                                   & \begin{tabular}[l]{@{}l@{}}  $L_{80}^4,L_{81}^4,L_{82}^4,L_{83}^4$ \end{tabular} \\
	    \hline\hline
	\end{tabular}
\end{table}

For each restoration stage, the restored loads and frequency nadir in MG1-MG4 are shown in Table \ref{Summary1}. Total 1773 kW of load are restored at the end of the 5 stages. It can be observed the service restoration actions happened in certain stages rather than in all stages. For example, MG1 restores 280.5 kW of load in Stage 1, but it restores no more load until Stage 4. While MG4 takes action on service restoration in each stage. It is because the sequential service restoration is limited by operational constraints, among which the maximum load step in each stage is again limited by the proposed frequency-dynamics constraints. Note that a larger amount of restored load in the optimization level will typically cause a lower frequency nadir in the simulation level, then a low frequency nadir will be considered in constraint \eqref{eq3_MSL} and help the optimization level to restrict a larger amount of restored load in next restoration stage. Because the first stage is the entry point of the restoration process, there is no prior frequency nadir information to be used in constraint \eqref{eq3_MSL}, therefore, the restored load in the first stage is typically the largest among all stages, which leads to a corresponding lowest frequency nadir among all stages. 
\begin{table}[]
	\centering
	\renewcommand{\arraystretch}{1.3}	
	\caption{Restored Loads, Frequency Nadir and Computation Time for MG1-MG4.}
	\label{Summary1}
	\begin{tabular}{clll}
		\hline\hline
		\multicolumn{2}{c}{Cases}                                                               
		& \begin{tabular}[c]{@{}l@{}}Restored load\\  (kW)\end{tabular} 
		& \begin{tabular}[c]{@{}l@{}}Frequency nadir\\ (Hz)\end{tabular} \\ \hline
		\multirow{5}{*}{\begin{tabular}[c]{@{}c@{}}MG1\\  ($T=4$ and $\alpha=0.1$)\end{tabular}} 	
		& Stage 1 & 280.5                                                  & 59.7044                                                     \\  
		& Stage 2 & 280.5                                                  & 59.9992                                                     \\ 
		& Stage 3 & 280.5                                                  & 59.9992                                                     \\ 
		& Stage 4 & 346.5                                                  & 59.9200                                                     \\		
		& Stage 5 & 346.5                                                  & 59.9989                                                     \\ 
		\hline	
		\multirow{5}{*}{\begin{tabular}[c]{@{}c@{}}MG2\\  ($T=4$ and $\alpha=0.1$)\end{tabular}}  		
		& Stage 1 & 230.0                                                  & 59.7079                                                     \\  
		& Stage 2 & 360.0\                                                 & 59.8201                                                     \\ 
		& Stage 3 & 420.0                                                  & 59.9146                                                     \\ 
		& Stage 4 & 420.0                                                  & 59.9984                                                     \\		
		& Stage 5 & 420.0                                                  & 59.9984                                                     	\\ 
		\hline	
		\multirow{5}{*}{\begin{tabular}[c]{@{}c@{}}MG3\\  ($T=4$ and $\alpha=0.1$)\end{tabular}} 		
		& Stage 1 & 212.5\                                                 & 59.7116                                                     \\  
		& Stage 2 & 212.5\                                                 & 59.9990                                                     \\ 
		& Stage 3 & 212.5\                                                 & 59.9990                                                     \\ 
		& Stage 4 & 382.5                                                  & 59.7656                                                     \\		
		& Stage 5 & 382.5                                                  & 59.9985                                                     	\\ 
		\hline		                                                
		\multirow{5}{*}{\begin{tabular}[c]{@{}c@{}}MG4\\  ($T=4$ and $\alpha=0.1$)\end{tabular}}      
		& Stage 1 & 192.0                                                  & 59.7910                                                     \\  
		& Stage 2 & 324.0                                                  & 59.8541                                                     \\ 
		& Stage 3 & 414.0                                                  & 59.9003                                                     \\ 
		& Stage 4 & 570.0\                                                 & 59.8230                                                     \\		
		& Stage 5 & 624.0                                                  & 59.9364                                                     	\\ \hline\hline
	\end{tabular}
\end{table}

\begin{figure} 
	\vspace{-0pt} 
	\vspace{-0pt}
	\centering
	\includegraphics[width=1.0\linewidth]{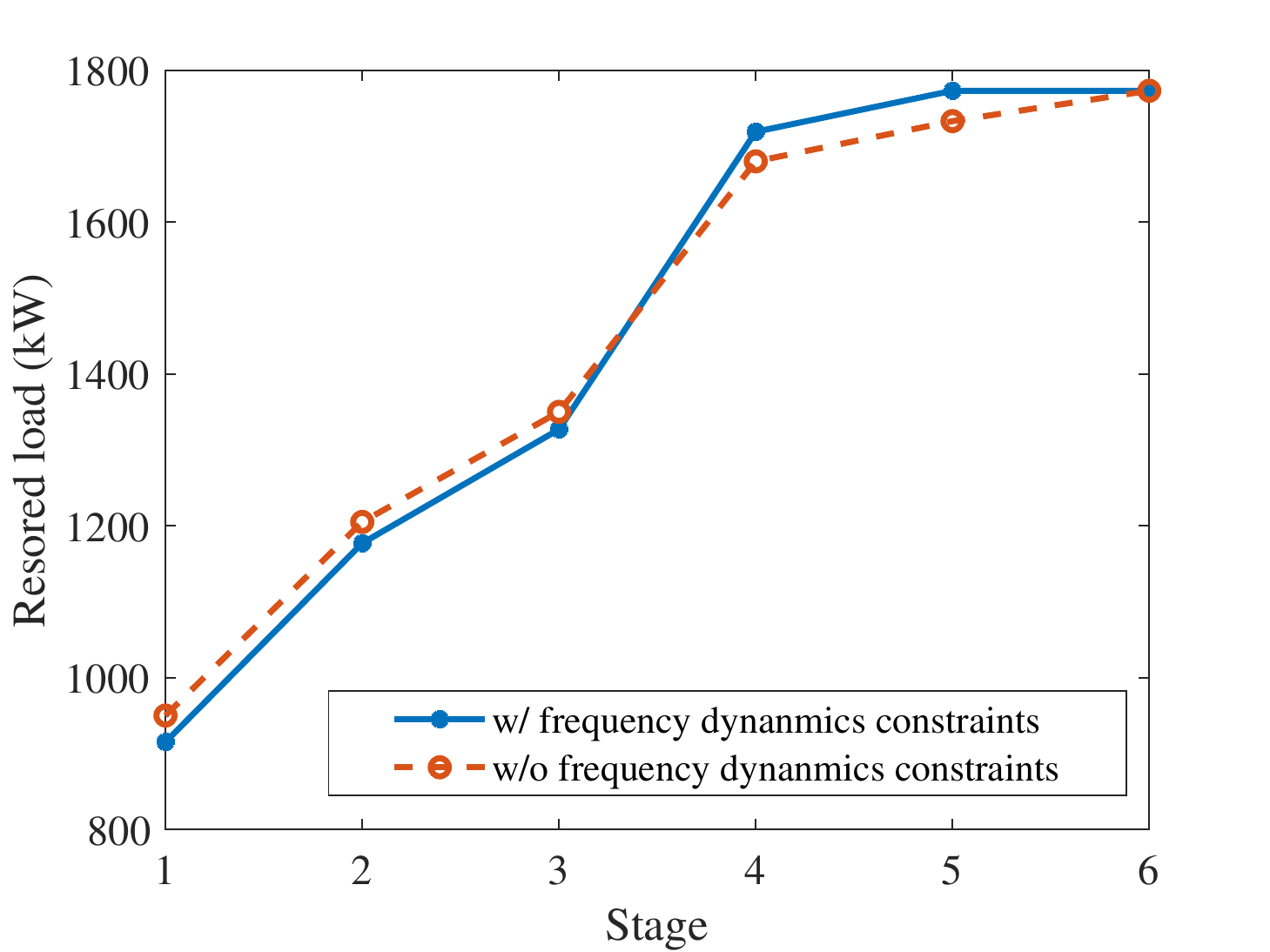}
	\vspace{-0pt}\
	\caption{Total restored load with and without considering frequency dynamics constraints.}  
	\centering
	\label{w_wo_alpha}
		\vspace{-0pt}\
\end{figure}

The comparison of total restored loads with and without considering the proposed frequency dynamics constraints is shown in Fig. \ref{w_wo_alpha}. Note that the total amount of restorable load of the base case model (i.e., without the frequency dynamics constraints) is the same as that of the proposed model with the frequency dynamics constraints. That is because the total load of the test system is fixed and less than the total DG generation capacity in both models. However, the base case needs 6 stages to fully restore the all the loads, while the proposed model can achieve that goal in the first 5 stages (as it is observed, no more loads between Stage 5 and Stage 6 are restored). While  In the early stages 1 to 3, the restored load of the proposed model is a little bit less than the base case. A further analysis is that: during the early restoration stages, the proposed model generated a restoration solution that prevents too low frequency nadir during transients. The base case restores more loads at Stage 1 to Stage 3 without considering such limitation on the frequency nadir. However, Stage 4 is a turning point when the proposed model restores more loads than the base case.  Therefore, the proposed model restores less loads than the base case during early stages (here, Stage 1 to Stage 3), while it restores more loads than the base case during later stages (from Stage 4). Such restoration pattern (restored load at each stage) of the base case model and the proposed model may vary case by case if the system topology or other operational constraints are changed. Therefore, if we implement the base case model and the proposed model in another test system with different topology or constraint settings, the base case model may restore fewer loads than the proposed model in the early stages and the turning point stage may change as well.           

\begin{figure} 
	\vspace{-0pt} 
	\vspace{-0pt}
	\centering
	\includegraphics[width=1.0\linewidth]{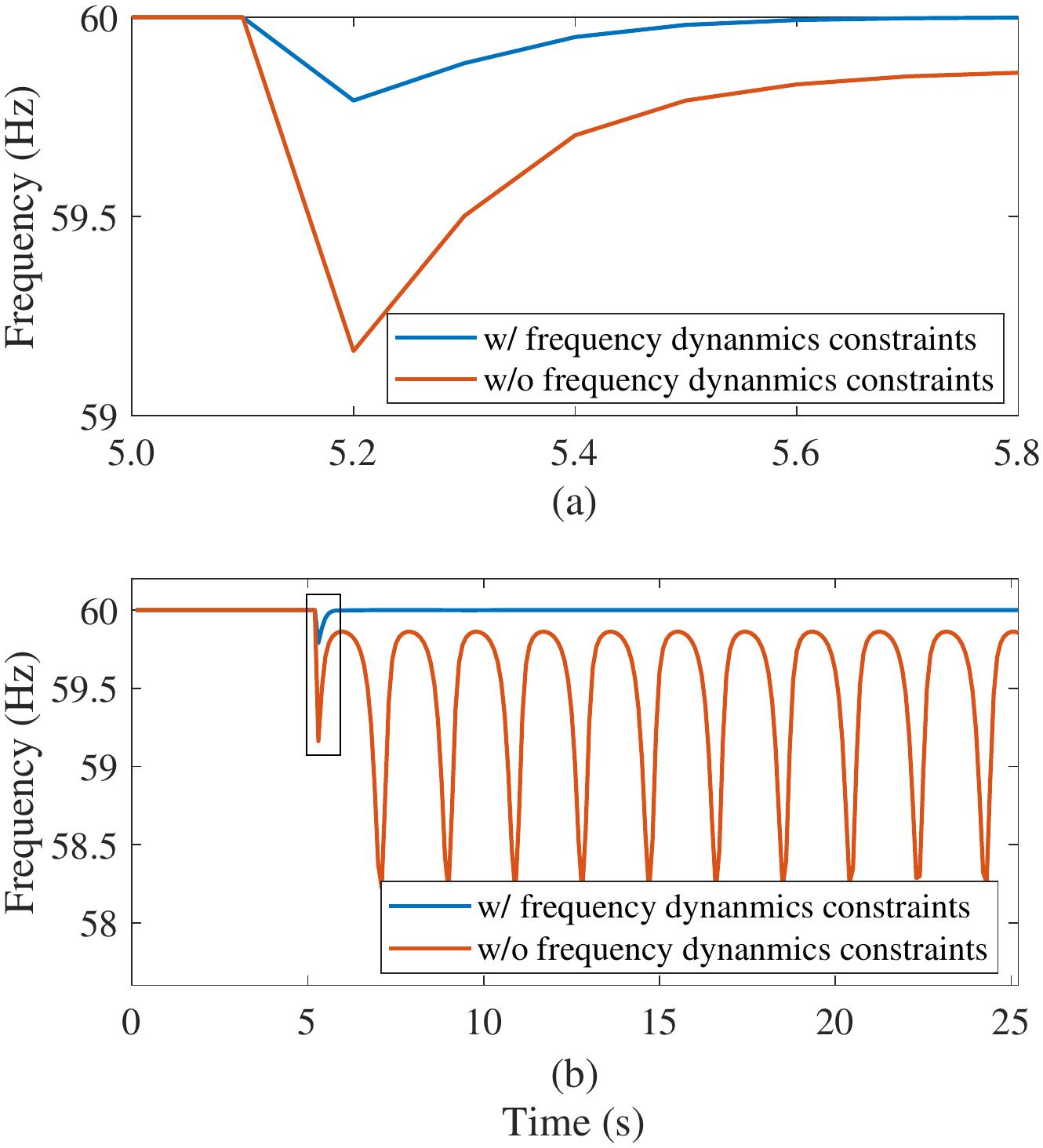}
	\vspace{-0pt}\
	\caption{Frequency responses of MG4 with and without frequency dynamics constraints: (a) Subplot of frequency response of MG4 during 5.0 s to 5.8 s; (b) Frequency responses of MG4 in Stage 1.}
	\centering
	\label{freq_w_wo_alpha}
		\vspace{-0pt}\
\end{figure}

In Fig. \ref{freq_w_wo_alpha}a and Fig. \ref{freq_w_wo_alpha}b, a zoom in view of the frequency response of MG4 and the frequency response of MG4 in Stage 1 are shown for better observation of the frequency dynamic performance. The frequency responses with and without the frequency dynamics constraints are represented by blue and red lines, respectively. By this comparison, it can be observed that both the rate of change of frequency and frequency nadir are significantly improved by considering frequency dynamics constraints in the proposed restoration model. However, if the frequency dynamics constraints are not considered to prevent a large frequency drop, unstable frequency oscillation may happen. The reason of the oscillation phenomenon in Fig. \ref{freq_w_wo_alpha}b is the too large $P^L$, which deviates the initial state of MG in the current stage out of the region of attraction of the original stable equilibrium. This in turn demonstrates the necessity to incorporate that frequency dynamics constraint in the optimization level. Note that $\omega_{\rm set}$ is set to 60 Hz in the droop equation (34), the equilibrium can be achieved when $\omega=\omega_{\rm set}$ and $P=P^L$, which means that the output frequency tracks the frequency reference when the output power of the simulation level tracks the target restored load calculated from the optimization level.   

\begin{figure} 
	\vspace{-0pt} 
	\vspace{-0pt}
	\centering
	\includegraphics[width=1.0\linewidth]{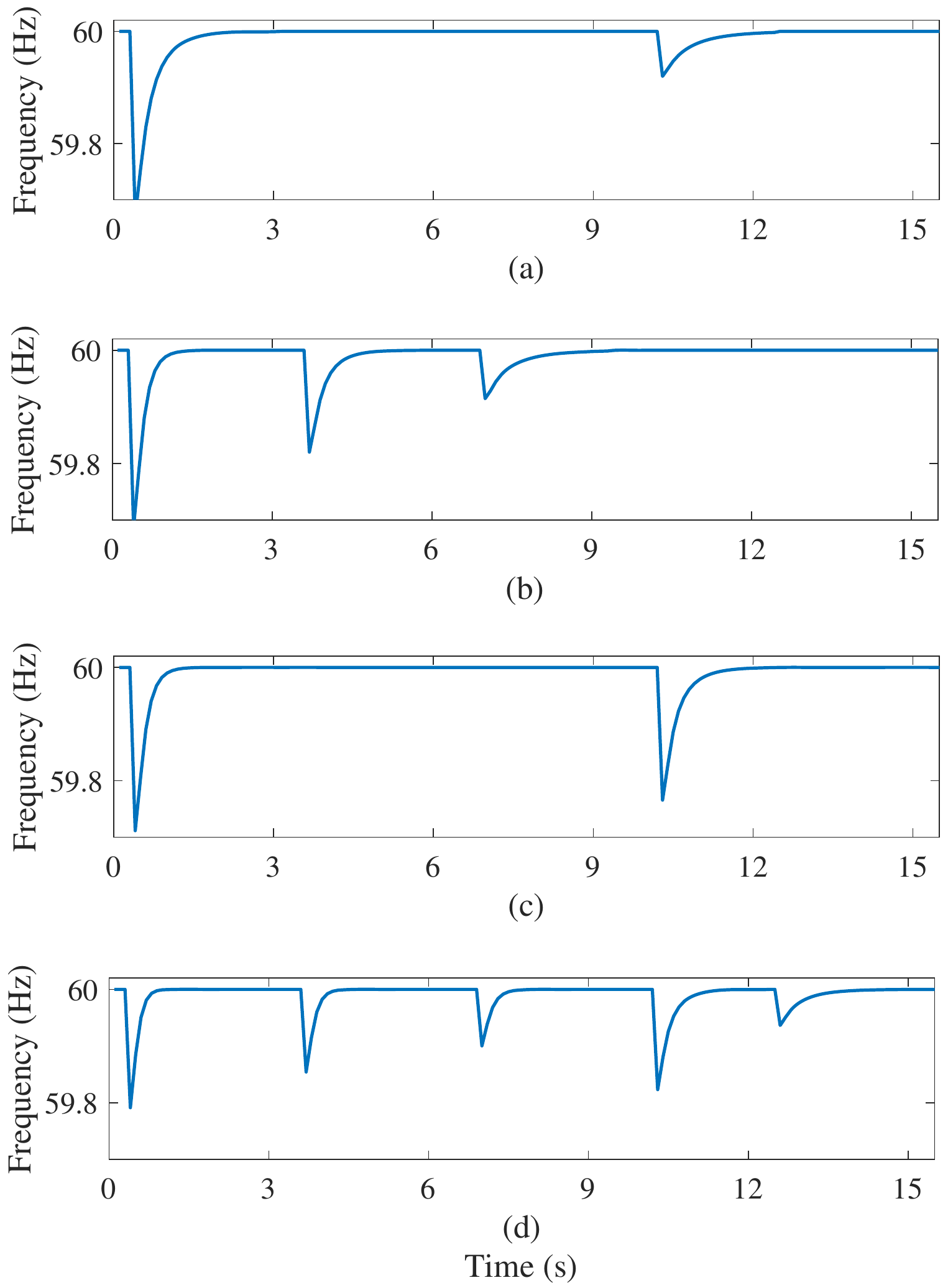}
	\vspace{-0pt}\
	\caption{Frequency responses of inverter-dominated MGs: (a) MG1; (b) MG2; (c) MG3; (d) MG4.}
	\centering
	\label{Freq_stage}
		\vspace{-0pt}\
\end{figure}

Fig. \ref{Freq_stage} shows the frequency responses of each inverter-dominated MG based on the proposed restoration model. The results show that the MG frequency drops when the load is restored. Because the maximum load step is constrained in the proposed MILP-based sequential service restoration model, the frequency nadir is also constrained. When load is restored as the frequency drops, the frequency nadir can be effectively maintained above the $f^{\rm min}$ threshold. 

\subsection{Impact of Hyper-parameters in Frequency Dynamics Constraints}
\begin{figure} 
	\vspace{-0pt} 
	\vspace{-0pt}
	\centering
	\includegraphics[width=1.0\linewidth]{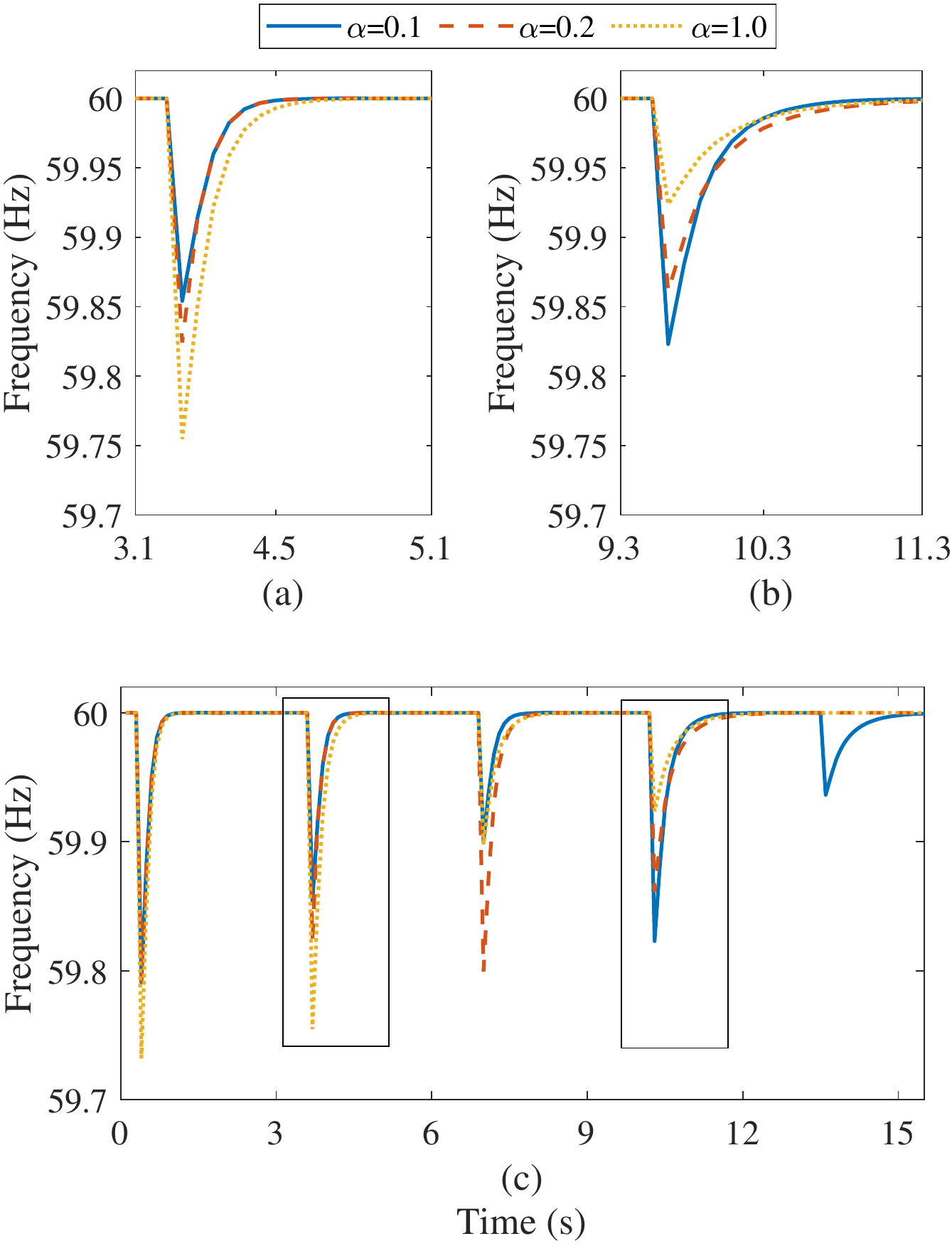}
	\vspace{-0pt}\
	\caption{Frequency responses of MG4 with different $\alpha$: (a) Frequency responses during 3.1 s to 5.1 s; (b) Frequency responses during 9.3 s to 11.3 s; (c) Frequency responses during the whole restoration process.}  
	\centering
	\label{diff_alpha}
		\vspace{-0pt}\
\end{figure}

Compared to other MGs, MG4 is heavily loaded with the largest number of nodes. Based on the results of Fig. \ref{results_123}, MG4 needs more stages to be fully restored compared to other MGs. Therefore, MG4 is chosen to test the effect of different $\alpha$ values. In Fig. \ref{diff_alpha}a and Fig. \ref{diff_alpha}b, the frequency responses of MG4 during the period of 3.1 s to 5.1 s, the period of 9.3 s to 11.3 s and the whole restoration process are shown, where the frequency with $\alpha=0.1$, $\alpha=0.2$ and $\alpha=1.0$ are represented by blue solid line, red dashed line and yellow dotted line, respectively. It can be observed that 5 stages are required to fully restore all the loads when $\alpha=0.1$; while only 4 restoration stages are needed when $\alpha=0.2$ or $\alpha=1.0$. During the period of 3.1 s to 5.1 s in left of Fig. \ref{diff_alpha}a, the frequency nadirs with $\alpha=0.2$ or $\alpha=1.0$ are lower than the frequency nadir with $\alpha=0.1$, which means more loads can be restored with larger value of $\alpha$. During the period of 9.3 s to 11.3 s in right of Fig. \ref{diff_alpha}b, the frequency nadir with $\alpha=0.1$ is lower than the frequency nadirs with $\alpha=0.2$ and $\alpha=1.0$, it is because the total restored loads for different $\alpha$ values are same, with $\alpha=0.2$ or $\alpha=1.0$, it can restore more loads in the early restoration stage, therefore they just need less loads to be restored in the late restoration stage. However,  $\alpha=0.1$ restores less loads in the early restoration stage, it has to restore more loads in the late restoration stage. As shown in Fig. \ref{diff_alpha}c, the overall dynamic frequency performance with $\alpha=0.1$ is still better than the cases with $\alpha=0.2$ and $\alpha=1.0$. Hence, there is a trade-off between dynamic frequency performance and restoration performance regarding the choice of $\alpha$: too small $\alpha$ may lead to too slow restoration and the frequency nadir may be high in the early restoration stage and the frequency nadir may be low in the late restoration stage; in turn, a large $\alpha$ may lead to less number of restoration stages, too large $\alpha$ may cause too low frequency in early stages and deteriorate the dynamic performance of the system frequency in a practical restoration process.  

We also shows that different values of the horizon length $T$ may cause different service restoration results. Table \ref{Summary2} summarizes the total restored loads and computation time using different horizon lengths in the proposed service restoration model. On the one side, the restored loads of case with $T=2$ and $T=3$ are less than that of the cases with $T \geq 4$, where the total restored load can reach the maximum level. Therefore, the results with small number of horizon length $T=2$ and $T=3$ are sub-optimal restoration solutions. On the other side, the longer horizon length also leads to heavy computation burden and increase the computation time. Similar to the impact of $\alpha$, there can be a trade-off between the computation time and the quality of solution when determining the value of $T$. 
\begin{table}[]
		\centering
		\renewcommand{\arraystretch}{1.3}		
		\caption{Restored Loads, Frequency Nadir and Computation Time with Different Horizon Lengths.}
		\label{Summary2}
			\vspace{-0pt}\
\begin{tabular}{cll}
\hline\hline
                        & Total restored load (kW) & Computation time (s) \\\hline
$T=2$                   & 1362.5  & 26.8870      \\
$T=3$                   & 1410.5  & 32.6725       \\
$T=4$                   & 1773.0  & 48.5629       \\
$T=5$                   & 1773.0  & 61.9968       \\
$T=6$                   & 1773.0  & 88.0216       \\
\hline\hline
\end{tabular}
\end{table}

\begin{figure} 
	\vspace{-0pt} 
	\vspace{-0pt}
	\centering
	\includegraphics[width=1.0\linewidth]{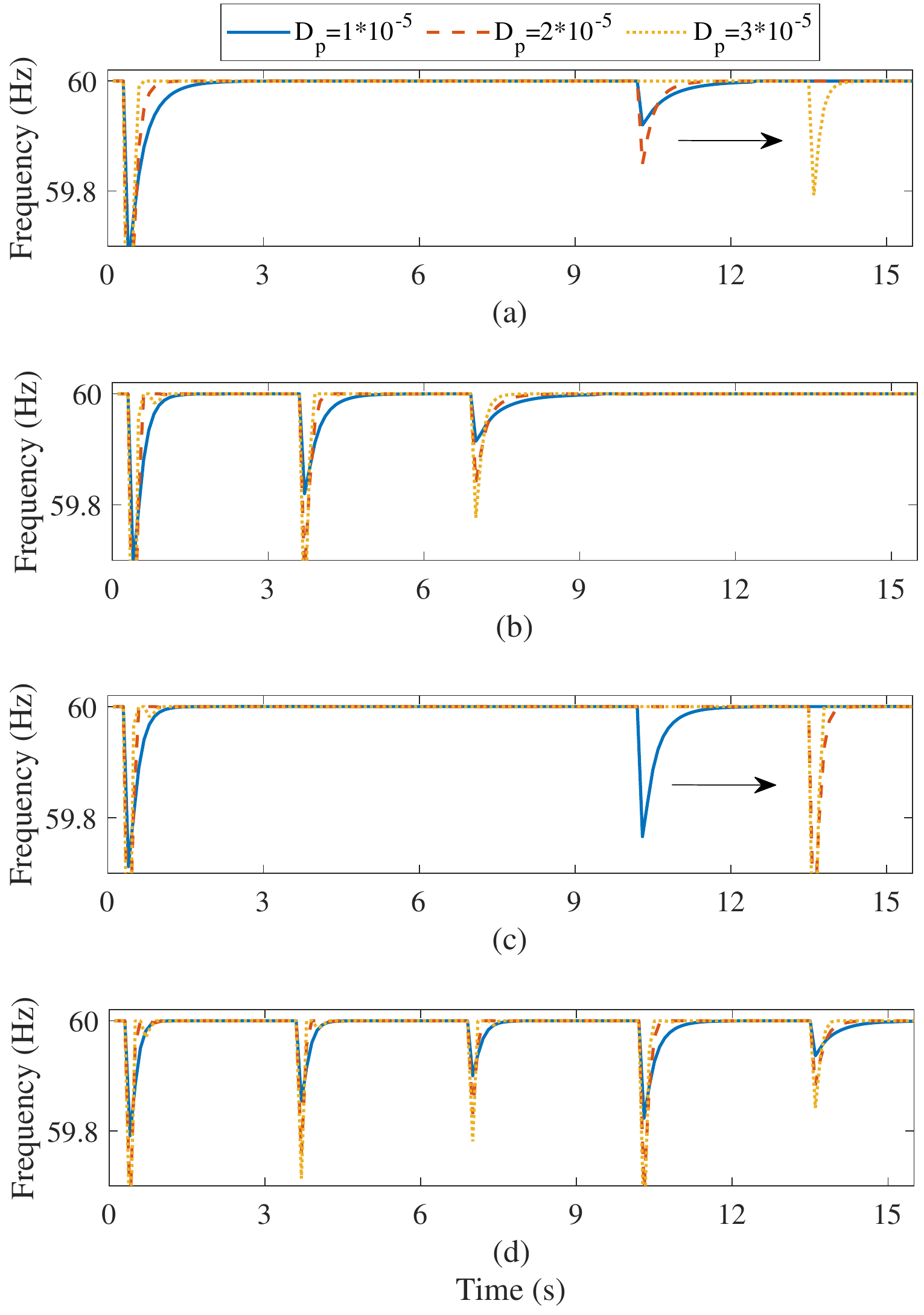}
	\vspace{-0pt}\
	\caption{Frequency responses of inverter-dominated MGs with different values of $D_p$ during restoration process: (a) MG1; (b) MG2; (c) MG3; (d) MG4.}  
	\centering
	\label{Freq_stage_diff_Dp}
		\vspace{-0pt}\
\end{figure}

In Fig. \ref{Freq_stage_diff_Dp}, the frequency responses of MG1 to MG4 are depicted during the restoration process with different values of droop gain $D_p$. In the test case, the original setting of $D_p$ is $1\times10^{-5}$. It can be observed that the different values of $D_p$ will cause different restoration solutions and frequency responses. As indicated by the arrow in Fig. \ref{Freq_stage_diff_Dp}a, MG1 can be fully restored in four stages when  $D_p=1\times10^{-5}$ or $2\times10^{-5}$, however, if the $D_p=3\times10^{-5}$, MG1 needs five stages to be fully restored. Similar observation can be found for restoration stage in Fig. \ref{Freq_stage_diff_Dp}c for MG3, it needs five stages to be fully restored when $D_p$ equals larger values (such as $2\times10^{-5}$ or $3\times10^{-5}$), while it only needs four stages when $D_p$ equals smaller values (such as $=1\times10^{-5}$). As shown in Fig. \ref{Freq_stage_diff_Dp}b and Fig. \ref{Freq_stage_diff_Dp}d, larger value of $D_p$ will also lead to larger frequency drop during restoration process.

\section{Conclusion}\label{sec:Con}
To improve the dynamic performance of the system frequency during service restoration of a unbalanced distribution systems in an inverter-dominated environment, we propose a simulation-assisted optimization model considering frequency dynamics constraints with clear physical meanings. Results demonstrate that: (i) The proposed frequency dynamics constrained service restoration model can significantly reduce the transient frequency drop during MGs forming and service restoration. (ii) Other steady-state performance indicators of our proposed method can rival that of the conventional methods, in terms of the final restored total load and the required number of restoration stages. Investigating on how to choose the best hyper-parameters, such as $\alpha$, horizon length $T$ and droop gain $D_p$ will be the next research direction.






\ifCLASSOPTIONcaptionsoff
  \newpage
\fi



\bibliographystyle{IEEEtran}
\bibliography{IEEEabrv,./bibtex/bib/IEEEexample}

\begin{IEEEbiography}[{\includegraphics[width=1in,height=1.25in,clip,keepaspectratio]{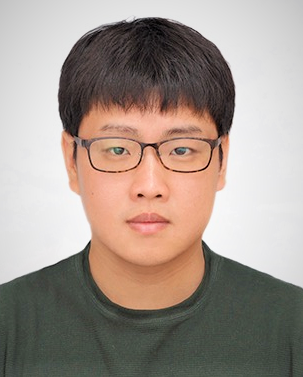}}]{Qianzhi Zhang}(S'17) is currently pursuing his Ph.D. in the Department of Electrical and Computer Engineering, Iowa State University, Ames, IA. He received his M.S. in electrical and computer engineering from Arizona State University in 2015. He has worked with Huadian Electric Power Research Institute from 2015 to 2016 as a research engineer. His research interests include the applications of machine learning and advanced optimization techniques in power system operation and control.
\end{IEEEbiography}
	
\begin{IEEEbiography}[{\includegraphics[width=1in,height=1.25in,clip,keepaspectratio]{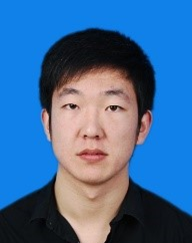}}]{Zixiao Ma}(S'18) is currently a Ph.D. student in the Department of Electrical and Computer Engineering at the Iowa State University, Ames, IA, USA. He received his B.S. degree in Automation and M.S. degree in Control theory and Control Engineering from Northeastern University in 2014 and 2017 respectively. His research interests are focused on the power system load modeling, microgrids, nonlinear control and model reduction.
\end{IEEEbiography}

\begin{IEEEbiography}[{\includegraphics[width=1in,height=1.25in,clip,keepaspectratio]{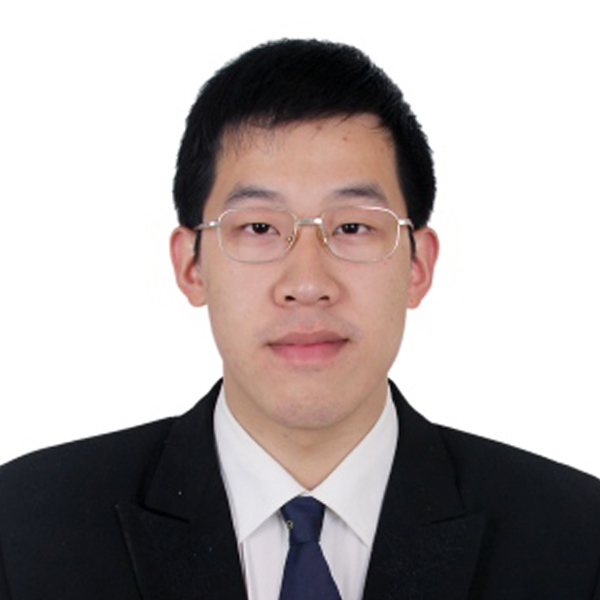}}]{Yongli Zhu}(S’12) received his B.S. degree from Huazhong University of Science and Technology in 2009, M.S. degree from State Grid Electric Power Research Institute in 2012, and Ph.D. degree from the University of Tennessee, Knoxville in 2018. He joined Iowa State University in the position of postdoc researcher in 2020. His research interests include power system stability, microgrid, and machine learning applications in power systems.
\end{IEEEbiography}

\begin{IEEEbiography}[{\includegraphics[width=1in,height=1.25in,clip,keepaspectratio]{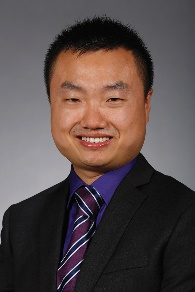}}]{Zhaoyu Wang}(S'13--M'15--SM’20) is the Harpole-Pentair Assistant Professor with Iowa State University. He received the B.S. and M.S. degrees in electrical engineering from Shanghai Jiaotong University, and the M.S. and Ph.D. degrees in electrical and computer engineering from Georgia Institute of Technology. His research interests include optimization and data analytics in power distribution systems and microgrids. He is the Principal Investigator for a multitude of projects focused on these topics and funded by the National Science Foundation, the Department of Energy, National Laboratories, PSERC, and Iowa Economic Development Authority. Dr. Wang is the Chair of IEEE Power and Energy Society (PES) PSOPE Award Subcommittee, Co-Vice Chair of PES Distribution System Operation and Planning Subcommittee, and Vice Chair of PES Task Force on Advances in Natural Disaster Mitigation Methods. He is an editor of IEEE Transactions on Power Systems, IEEE Transactions on Smart Grid, IEEE Open Access Journal of Power and Energy, IEEE Power Engineering Letters, and IET Smart Grid. Dr. Wang was the recipient of the National Science Foundation (NSF) CAREER Award, the IEEE PES Outstanding Young Engineer Award, and the Harpole-Pentair Young Faculty Award Endowment.
\end{IEEEbiography}

\end{document}